\documentclass[showpacs,showkeys,prc,aps,preprintnumbers]{revtex4}
\usepackage{graphicx}
\usepackage{verbatim}
 
\newcommand{ \be }{\begin{equation}}
\newcommand{ \ee }{\end{equation}}
\newcommand{ \bea }{\begin{eqnarray}}
\newcommand{ \eea }{\end{eqnarray}}

\usepackage{color}

\begin{document}

\title{Methods for the Study of Transverse Momentum Differential Correlations }

\author{Monika Sharma and Claude A. Pruneau}

\affiliation{Physics and Astronomy Department, Wayne State University, 
Detroit, MI 48201 USA}

\begin{abstract}
We introduce and compare three differential correlation functions for the study of transverse momentum  
correlation in $p+p$ and $A+A$ collisions.  These consist of {\it inclusive}, {\it event-wise} and a differential version of the 
correlation measure $\tilde C$ introduced by Gavin \cite{Gavin} for experimental study of the viscosity per unit 
entropy of the matter produced in $A+A$ collisions. We study the quantitative difference between the three 
observables on the basis of PYTHIA simulations of $p+p$ collisions and $A+A$ collisions consisting of an arbitrary  
superposition of $p+p$ collision events at $\sqrt{s} = $200 GeV. We observe that {\it inclusive} and {\it event-wise} correlation functions are 
remarkably identical to each other where as the observable $\tilde C$ differs from the two. 
We study the robustness and efficiency dependencies of these observables based on truncated Taylor expansions 
in efficiency in $p+p$ collisions and on the basis of Monte Carlo simulation using an adhoc detector efficiency parameterization. 
We find that all the three observables are essentially independent of detector efficiency. 
We additionally study the scaling of the correlation measures and find all the observables exhibit an 
approximate $1/N$ dependence of the number of participants ({\it N}) in $A+A$ collisions. Finally, we study the impact of 
flow-like anisotropy on the {\it inclusive} correlation function and find flow imparts azimuthal modulations similar to 
those observed with two-particle densities.  
\end{abstract}

\maketitle

\section{Introduction}
\label{Sect:Introduction}

Measurements of two- and multi-particle correlations have proven a powerful tool to study heavy ion collision 
dynamics. In particular, measurements of two-particle correlations in the form of two-particle densities as a 
function of the particle pseudorapidities and azimuthal angles in $A+A$ collisions have revealed new correlation 
features not found in $p+p$ interactions. Specifically, azimuthal two-particle correlations in central $A + A$ 
collisions exhibit novel ``away-side" structures, commonly referred as the away-side dip, and a near-side 
structure called the ``ridge". First measurements of the ``dip" and ``ridge" were reported on the basis of {\it inclusive} 
charged particle correlations \cite{dipRidge}.  More recently, measurements were extended to include identified 
charged and neutral particles, studies of correlations in various momentum ranges and system size dependencies 
\cite{SoftRidge}. Measurements of event-by-event fluctuation observables such as net charge and transverse 
momentum provide complementary information. Studies of net charge and transverse momentum fluctuations 
show that while the magnitude  of the fluctuations observed in $A+A$ collisions differs from that observed in 
$p+p$ collisions (after proper scaling to account for the number of collision participants), no dramatic 
suppression of the net charge fluctuations is observed that might signal the formation of a Quark Gluon Plasma (QGP) 
\cite{Pruneau03,Pruneau08}. Similarly, only small differences are observed in the magnitude of $p_t$ dynamical 
fluctuations in $A+A$ collisions relative to those observed in $p+p$ interactions after scaling by the number of 
participants and the particle average $p_t$.  The measured dynamical $p_t$ fluctuations in $A+A$ thus do not provide 
evidence for the strongly enhanced fluctuations expected for nuclear matter near the tri-critical point \cite{StarPtPt}. 
In contrast, however, the width of the charge balance function exhibits a significant reduction from peripheral to 
central $Au+Au$ collisions that is consistent with the predictions of delayed hadronization when a QGP is 
formed \cite{StarBalanceFct}. The reduction of the balance function measured as a function of pseudorapidity may, 
however, result from radial flow, resonance decays, or other effects. It is thus important to seek 
additional observables to obtain a better understanding of the $A+A$ collision dynamics. 

In this work, we consider measurements of differential transverse momentum two-particle correlations as a 
function of the relative pseudorapidity and azimuthal angles of measured particles.  A certain latitude exists in the 
calculation of such correlation functions. Indeed, as for integral correlations \cite{StarPtPt,Adams:2004gp}, one 
can use both {\it inclusive} and {\it event-wise} definitions. One can also consider dynamical  fluctuations as the difference 
between measured fluctuations and statistical fluctuations i.e. those expected for a purely Poisson system. 
Extensions of fluctuation variables $\Phi_{p_{t}}$ \cite{GazdzickiMrowczynski} and $\sigma_{p_{t}}$ \cite{Trainor} are 
also possible. In this work, we study three differential observables namely, {\it inclusive} and {\it event-wise} correlation 
functions and a differential version of the measure $\tilde C$, used by Gavin \cite{GavinAbdelAziz07,Gavin08} 
to estimate the viscosity per unit entropy of the matter formed in $A+A$ collisions. 
The observable definitions and notation used 
in this work are presented in Sect. \ref{Sect:Definition}. In Sect. \ref{Sect:PtCorrelations}, we focus on the 
definition of these observables for the study of transverse momentum correlations between two particles as 
functions of the relative pseudorapidity, and azimuthal angles of the particles. The goal of these observables is 
essentially to measure the covariance of two particle momenta by averaging the product of the differences 
between  these momenta and the mean momentum. Given their different definition, one expects, in practice, they 
should measure  different numerical values. The quantitative differences between these observables are assessed in Sect. 
\ref{Sect:Comparison} on the basis of generic arguments and PYTHIA simulations. Section \ref{Sect:Model} 
presents predictions based on the event generator PYTHIA \cite{Pythia}. Sect. \ref{Sect:Scaling} discusses the dependence 
of these observables on the number of participants in $A+A$ collisions.  Elliptic flow effects on the magnitude 
and azimuthal dependence of these observables is discussed in Sect. \ref{Sect:FlowEffects}. The experimental 
robustness of these observables is studied in Sect. \ref{Sect:Robustness}. 
This work is summarized in Sect. \ref{Sect:Summary}.

\section{Notation and Definitions}
\label{Sect:Definition}

We consider measurements of one- and two-particle production cross-sections. We are specifically interested 
in the measurements of two-particle correlations. Let $x_i $ and $y_i $
represent arbitrary observables for two particles, {\it i}=1,~2. These observables may be the momenta, azimuthal 
angle, pseudorapidity etc., of the particles. Let $\rho _1 \left( {x_1 } \right)$
 and $\rho _2 \left( {x_1 ,x_2 } \right)$ be one- and two-particle densities, respectively, defined as 
functions of these observables.

\begin{eqnarray}
\rho _1 \left( {x,y} \right) & = & \frac{{d N_1}}{{dxdy}} \\ 
\label{Eq1}
\rho _2 \left( {x_1 ,x_2 ,y_1 ,y_2 } \right) & = & \frac{{d N_2 }}{{dx_1 dx_2 dy_1 dy_2 }}
\label{Eq2}
\end{eqnarray}

Discussions in this paper are limited to two-particle correlations but can be straightforwardly extended to higher 
order multi-particle correlations. For simplicity, and illustrative purposes, assume the particles are 
indistinguishable. Integrals of the above densities over the space spanned by observables ($x_{i},y_{i}$) yield the 
number of particles and pairs of particles, respectively, in the domain of integration.
\begin{equation}
\int{\rho_1\left(x,y\right)dxdy} = \int{\frac{{d N_1}}{{dxdy}}dxdy} = N 
\end{equation}
\begin{equation}
\int {\rho _2 \left( {x_1 ,x_2 ,y_1 ,y_2 } \right)dx_1 dx_2 dy_1 dy_2 }  = \int {\frac{{dN_2 }}{{dx_1 
dx_2 dy_1 dy_2 }}dx_1 dx_2 dy_1 dy_2 }  = N\left( {N - 1} \right)
\label{Eq4}
\end{equation}
	
where {\it N} is the average number of particles produced in the region of acceptance of interest. 

Experimentally, the number of particles in a given acceptance fluctuates event-by-event owing to the stochastic 
nature of the production processes. The one- and two-particle densities may then be viewed as an average of the 
number of particles taken over all measured events.  We label events with an index $\alpha=1, ..., N_{ev}$, 
where $N_{ev}$ is the number of events in the data sample. One measures the number of particles, $n_\alpha(y)$ at a 
given location $y$ in a bin of width $\Delta y$ for each event $\alpha$. The one- particle and two- particle 
densities are then obtained by averaging over the event ensemble, i.e. the total number of events $N_{ev}$.
\begin{equation}
\rho _1 \left( y \right) \equiv \int {\rho _1 \left( {x,y} \right)dx} 
= \frac{1}{{\Delta y}}\left\langle {n(y)} \right\rangle  
= \frac{1}{{\Delta y}}\frac{1}{{N_{ev} }}\sum\limits_{\alpha  = 1}^{N_{ev} } {n_\alpha  (y)}
\label{Eq5}
\end{equation}
\begin{equation}
\rho _2 \left( {y_1 ,y_2 } \right) \equiv  \int {\rho _2 \left( {x_1 ,x_2 ,y_1 ,y_2 } \right)dx_1 dx_2 }
= \frac{1}{{\Delta y^2 }}\left\langle {n(y_1 )n(y_2 )} \right\rangle  
= \frac{1}{{\Delta y^2 }}\frac{1}{{N_{ev} }}\sum\limits_{\alpha  = 1}^{N_{ev} } {n_\alpha  (y_1 )n_\alpha  (y_2 )} 
\label{Eq6}
\end{equation}
where $\left\langle O \right\rangle $ represents the event-ensemble average of $O$.

In this work we describe measurements of fluctuations and a differential correlation observable. This requires 
measurements of covariances $\left\langle {\Delta x_1 \Delta x_2 } \right\rangle $ where $\Delta x = x - \left\langle x \right\rangle $ 
and $\left\langle x \right\rangle $ is the mean value of {\it x}, which is a function of coordinates $y_1$ and $y_2$. 
Such a function of $y_1$ and $y_2$ 
corresponds to a $\Delta x_1 \Delta x_2$ average taken with respect to the two-particle density given by Eq. 2. To denote 
such averages we use the notation $\rho _n^{g(x)} (y)$, where g(x) is some function of the coordinates $x_1$, $x_2$ and 
{\it n} = 1,2 signifies the one- or two- particle density, respectively. 
For example, the average value of {\it x} as a function of {\it y}, as defined below, denoted $\rho _1^x (y)$, is defined and 
calculated as:
\begin{equation}
\rho _1^x \left( y \right) \equiv \frac{{\int {\rho _1 \left( {x,y} \right)xdx} }}{{\int {\rho _1 \left( {x,y} \right)dx} }}
= \left\langle {x(y)} \right\rangle  = \frac{{\sum\limits_{\alpha  = 1}^{N_{ev} } {\sum\limits_{i =
1}^{n_\alpha  (y)} {x_{\alpha ,i} (y)} } }}{{\sum\limits_{\alpha  = 1}^{N_{ev} } {n_\alpha  (y)} }}
\label{Eq7}
\end{equation}
where $x_{\alpha ,i} (y)$ is the value of $x$ for particle $i$ at position $y$ in an event $\alpha$.
similarly the two-particle covariance $\left\langle {\Delta x_1 \Delta x_2 } \right\rangle $, a function of $y_1$ and $y_2$ 
is denoted $\rho _2^{\Delta x_1 \Delta x_2 } (y_1 ,y_2 )$ and is defined and calculated as:
\begin{eqnarray} 
\rho _{_2 }^{\Delta x_1 \Delta x_2 } \left( {y_1 ,y_2 } \right) \equiv \frac{{\int {\rho _2 \left( {x_1 
,x_2 ,y_1 ,y_2 } \right)\Delta x_1 \Delta x_2 dx_1 dx_2 } }}{{\int {\rho _2 \left( {x_1 ,x_2 ,y_1 ,y_2 } 
\right)dx_1 dx_2 } }} = \left\langle {\Delta x(y_1 )\Delta x(y_2 )} \right\rangle  \nonumber\\
= \frac{{\sum\limits_{\alpha  = 1}^{N_{ev} } {\sum\limits_{i = 1}^{n_\alpha  (y_1 )} {\sum\limits_{j \ne i =
1}^{n_\alpha  (y_2 )} {\left( {x_{\alpha,i}  (y_1 ) - \left\langle {x(y_1 )} \right\rangle } \right)\left( {x_{\alpha,j}  (y_2
) - \left\langle {x(y_2 )} \right\rangle } \right)} } } }}{{\sum\limits_{\alpha  = 1}^{N_{ev} } {n_\alpha  (y_1
)n_\alpha  (y_2 )} }} 
\label{Eq8}
\end{eqnarray}

The number of pairs in an event $\alpha$ is given by the product of the number of particles, $n_\alpha (y_1 )n_\alpha  (y_2 )$, 
in bins of width $\Delta y$ at positions $y_1$ and $y_2$. 
However, the number of pairs is reduced to $n_\alpha (y_1 )\left( {n_\alpha  (y_1 ) - 1} \right)$ for identical bins $y_1=y_2$ 
(indistinguishable particles).

The above expressions correspond to what is commonly referred to as {\it inclusive} averages.  An alternative 
averaging method is also available which corresponds to {\it event-wise} average. {\it event-wise} 
averages of $x$ and $\Delta x_1 \Delta x_2$ are herein noted as $\tilde \rho _1^x (y)$
and $\tilde \rho _2^{\Delta x_1 \Delta x_2 } (y_1 ,y_2 )$,
 respectively. They are calculated as follows:
\begin{eqnarray}
\tilde \rho _{_1 }^x \left( y \right) &=& \frac{1}{{N_{ev} }}\sum\limits_{\alpha  = 1}^{N_{ev} }                                                    
{\frac{1}{{n_\alpha  (y)}}\sum\limits_{i = 1}^{n_\alpha  (y)} {x_{\alpha ,i} (y)} }  \\ 
\label{Eq9}
\tilde \rho _{_2 }^{\Delta x_1 \Delta x_2 } \left( {y_1 ,y_2 } \right)
&=& \frac{1}{{N_{ev} }}\sum\limits_{\alpha = 1}^{N_{ev} } {\frac{1}{{n_\alpha  (y_1 )}}\frac{1}{{n_\alpha  (y_2 )}}}A 
\label{Eq10}
\end{eqnarray}
where
\begin{equation}
A = \sum\limits_{i = 1}^{n_\alpha  (y_1
)} {\sum\limits_{j \ne i = 1}^{n_\alpha  (y_2 )} {\left( {x_{\alpha ,i} (y_1 ) - \left\langle {x(y_1 )} \right\rangle
} \right)\left( {x_{\alpha ,j} (y_2 ) - \left\langle {x(y_2 )} \right\rangle } \right)} } 
\end{equation}
For integral correlations where the above averages are taken over a single wide $y$ bin, the above expression 
(Eq.10) reduces to fluctuation variables reported in recent works \cite{Voloshin05,StarPtPt}.

Gavin suggested recently \cite{Gavin08} that the width of transverse momentum fluctuation distributions 
measured as a function of pseudorapidity may be used to estimate the viscosity of the matter 
formed in $A+A$ collisions studied at RHIC and LHC. 

Here we consider a differential version of the correlation $\tilde C$, and write
\begin{equation}
\tilde C = \frac{{\sum\limits_{\alpha  = 1}^{N_{ev} } {\sum\limits_{i = 1}^{n_\alpha  \left( {y_1 } \right)} {\sum\limits_{i \ne j = 1}^{n_\alpha  \left( {y_2 } \right)} {x_{\alpha ,i} \left( {y_1 } \right)x_{\alpha ,j} \left( {y_2 } \right)} } } }}{{\sum\limits_{\alpha  = 1}^{N_{ev} } {n_\alpha  \left( {y_1 } \right)} n_\alpha  \left( {y_2 } \right)}} - \left( {\frac{{\sum\limits_{\alpha  = 1}^{N_{ev} } {\sum\limits_{i = 1}^{n_\alpha  \left( {y_1 } \right)} {x_{\alpha ,i} \left( {y_1 } \right)} } }}{{\sum\limits_{\alpha  = 1}^{N_{ev} } {n_\alpha  \left( {y_1 } \right)} }}} \right)\left( {\frac{{\sum\limits_{\alpha  = 1}^{N_{ev} } {\sum\limits_{j = 1}^{n_\alpha  \left( {y_2 } \right)} {x_{\alpha ,j} \left( {y_2 } \right)} } }}{{\sum\limits_{\alpha  = 1}^{N_{ev} } {n_\alpha  \left( {y_2 } \right)} }}} \right)
\label{Eq11}
\end{equation}
A number of other observables have been used recently to carry out measurements of fluctuations and are well documented in the 
literature. The quantity $\Phi _{p_t } $, defined in \cite{GazdzickiMrowczynski}, 
was used by CERN experiments NA49 and CERES to measure 
transverse momentum fluctuations in Pb + Pb collisions.  The quantity $\sigma _{p_{t},dyn}^2$
 was used by STAR to measure fluctuations in Au + Au collisions \cite{Voloshin:2005qj}. There is an approximate 
equivalence between these different integral observables as shown in \cite{VoloshinPRD60_1999}

\section{Transverse Momentum Differential Correlations}
\label{Sect:PtCorrelations}

In this section, we focus our discussion on integral and differential transverse momentum correlations. Integral 
correlations, $\left\langle {\Delta p_t \Delta p_t } \right\rangle $, 
may in principle provide an estimate of event-by-event temperature fluctuations which, in turn, may provide means 
to determine the heat capacity of the medium formed in $A+A$ collisions 
\cite{Stodolsky95,KorusMrowczynski2001,Stephanov0110077}. Additionally, one can also use the measure $\tilde C$, defined in Sect. 2, 
to study the transverse momentum correlations \cite{SoftRidge,GavinAbdelAziz07,Gavin08}. We argue that a 
differential version of all the observables, as defined in the previous section, may provide an equivalent or even 
better way to estimate the medium viscosity. However, calculation of viscosity using these observables is not presented in this work. 
Clearly, differential correlation functions provide more information 
than integrals over a wide range of pseudorapidity. Therefore they may enable a deeper understanding of the 
collision dynamics at play in $A+A$ collisions. As a specific example one notes the discovery of the ``ridge" in 
two-particle (density) correlations measured in $A+A$ collisions.  It is thus interesting to consider differential 
transverse momentum correlations as a function of differences in particle pseudorapidity and azimuthal angles  
(denoted as $\Delta \eta $ and $\Delta \varphi $, respectively). 

To simplify the notation, we use $p$ rather than the ``usual" $p_t$ to denote transverse momentum. 
We define the transverse momentum correlations of interest by substituting $p$ for $x$ in Eq. \ref{Eq7} through Eq. \ref{Eq11}, 
and include a dependency on the particles' pseudorapidity, $\eta$, and azimuthal angle, $\varphi$. The 
{\it inclusive} momentum correlation function is defined as
\begin{equation}
\rho _{_2 }^{\Delta p_1 \Delta p_2 } \left( {\Delta \eta ,\Delta \varphi } \right) \equiv \frac{{\int {\rho 
_2 \left( {p_1 ,p_2 ,\Delta \eta ,\Delta \varphi } \right)\Delta p_1 \Delta p_2 dp_1 dp_2 } }}{{\int {\rho _2 \left( 
{p_1 ,p_2 ,\Delta \eta ,\Delta \varphi } \right)dp_1 dp_2 } }}
\label{Eq12}
\end{equation}
and calculated using the following expression:
\begin{eqnarray}
\rho _{_2 }^{\Delta p_1 \Delta p_2 } \left( {\Delta \eta ,\Delta \varphi } \right) = 
\frac{{\sum\limits_{\alpha  = 1}^{N_{ev} } {\sum\limits_{i = 1}^{n_\alpha  (\eta _1 ,\varphi _1 )} 
{\sum\limits_{j \ne i = 1}^{n_\alpha  (\eta _2 ,\varphi _2 )} {\left( {p_{\alpha ,i} (\eta _1 ,\varphi _1 ) - 
\left\langle {p(\eta _1 ,\varphi _1 )} \right\rangle } \right)\left( {p_{\alpha ,j} (\eta _2 ,\varphi _2 ) - \left\langle 
{p(\eta _2 ,\varphi _2 )} \right\rangle } \right)} } } }}{{\sum\limits_{\alpha  = 1}^{N_{ev} } {n_\alpha  (\eta _1 
,\varphi _1 )n_\alpha  (\eta _2 ,\varphi _2 )} }}
\label{Eq13}
\end{eqnarray}
where one applies the conditions $\Delta \eta  = \eta _1  - \eta _2 $ and $\Delta \varphi  = \varphi _1  - \varphi _2 $. 
$n_{\alpha}(\eta_i,\varphi_i)$ is the number of particles detected in a given event $\alpha$ at pseudorapidity 
$\eta_i$ and laboratory angle $\varphi_i$. $p_{\alpha,i}(\eta_i,\varphi_i)$ represents the transverse momentum 
of particle $i$ from event $\alpha$. $\left\langle {p(\eta _1 ,\varphi _1 )} \right\rangle $
is the {\it inclusive} average particle transverse momentum at $\eta_i$ and $\varphi_i$. 
\begin{equation}
\left\langle {p\left( {\eta ,\varphi } \right)} \right\rangle  \equiv \rho _1^p \left( {\eta ,\varphi } \right)
= \frac{{\sum\limits_{\alpha  = 1}^{N_{ev} } {\sum\limits_{i = 1}^{n_\alpha  (\eta ,\varphi )} {p_{\alpha ,i} 
(\eta,\varphi )} } }}{{\sum\limits_{\alpha  = 1}^{N_{ev} } {n_\alpha  (\eta ,\varphi )} }}
\label{Eq14}
\end{equation}
We include dependencies of the average transverse momentum, $p$, on the laboratory angles $\varphi_1$ and 
$\varphi_2$ to account for possible instrumental effects. The mean $p$ is independent of azimuthal angles $\varphi_1$ and 
$\varphi_2$ for unpolarized beams.  The detector acceptance and particle detection efficiency may, however, 
vary with these angles and must thus be accounted for in practice. 

The {\it event-wise} transverse momentum correlation is calculated as follows.
\begin{eqnarray}
\tilde \rho _2^{\Delta p_1 \Delta p_2 } \left( {\Delta \eta ,\Delta \varphi } \right) = \frac{1}{{N_{ev} }}\sum\limits_{\alpha  = 1}^{N_{ev} } {\frac{1}{{n_\alpha  \left( {\eta _1 ,\varphi _1 } \right)}}} \frac{1}{{n_\alpha  \left( {\eta _2 ,\varphi _2 } \right)}} \nonumber \\
\times \sum\limits_{i = 1}^{n_\alpha  \left( {\eta _1 ,\varphi _1 } \right)} {\sum\limits_{i \ne j = 1}^{n_\alpha  \left( {\eta _2 ,\varphi _2 } \right)} {\left( {p_{\alpha ,i} \left( {\eta _1 ,\varphi _1 } \right) - \tilde \rho _1^p \left( {\eta ,\varphi } \right)} \right)\left( {p_{\alpha ,j} \left( {\eta _2 ,\varphi _2 } \right) - \tilde \rho _1^p \left( {\eta ,\varphi } \right)} \right)} }
\label{Eq15}
\end{eqnarray}

and uses the {\it event-wise} average
\begin{equation}
\overline {p\left( {\eta ,\varphi } \right)}  \equiv \tilde \rho _{_1 }^p \left( {\eta ,\varphi } \right)
= \frac{1}{{N_{ev} }}\sum\limits_{\alpha  = 1}^{N_{ev} } {\frac{1}{{n_\alpha  (\eta ,\varphi )}}\sum\limits_{i 
= 1}^{n_\alpha  (\eta ,\varphi )} {p_{\alpha ,i} (\eta ,\varphi )} } 
\label{Eq16}
\end{equation}
Similarly $\tilde C$ can be written as:
\begin{equation}
\tilde C = \frac{{\sum\limits_{\alpha  = 1}^{N_{ev} } {\sum\limits_{i = 1}^{n_\alpha  \left( {\eta _1 ,\varphi _1 } \right)} {\sum\limits_{i \ne j = 1}^{n_\alpha  \left( {\eta _2 ,\varphi _2 } \right)} {p_{\alpha ,i} \left( {\eta _1 ,\varphi _1 } \right)p_{\alpha ,j} \left( {\eta _2 ,\varphi _2 } \right)} } } }}{{\sum\limits_{\alpha  = 1}^{N_{ev} } {n_\alpha  \left( {\eta _1 ,\varphi _1 } \right)n_\alpha  \left( {\eta _2 ,\varphi _2 } \right)} }} - \left( {\frac{{\sum\limits_{\alpha  = 1}^{N_{ev} } {\sum\limits_{i = 1}^{n_\alpha  \left( {\eta _1 ,\varphi _1 } \right)} {p_{\alpha ,i} \left( {\eta _1 ,\varphi _1 } \right)} } }}{{\sum\limits_{\alpha  = 1}^{N_{ev} } {n_\alpha  \left( {\eta _1 ,\varphi _1 } \right)} }}} \right)\left( {\frac{{\sum\limits_{\alpha  = 1}^{N_{ev} } {\sum\limits_{j = 1}^{n_\alpha  \left( {\eta _2 ,\varphi _2 } \right)} {p_{\alpha ,j} \left( {\eta _2 ,\varphi _2 } \right)} } }}{{\sum\limits_{\alpha  = 1}^{N_{ev} } {n_\alpha  \left( {\eta _2 ,\varphi _2 } \right)} }}} \right)
\end{equation}

\section{Comparison of Differential Correlation Observables}
\label{Sect:Comparison}

The {\it inclusive} and {\it event-wise} differential correlation observables, as well as the correlation measure $\tilde C$, are 
designed to study the magnitude of correlation in terms of some specific kinematic variable {\it x}, where {\it x} may 
be, {\it e.g.,} the particle transverse momentum or pseudorapidity. The three observables represent averages of the difference 
between the  {\it x} value of each particle relative to the mean value of {\it x}. The three observables, however, employ distinct 
averaging methods and in general lead to different numerical results.  None of these averaging methods can be 
considered ``superior" in any sense. However, due to practical or technical reasons, studies of 
correlations are typically reported using one type of averaging only. Still other averaging methods were 
advocated and used in recent works \cite{Phenix,Voloshin2002,CERES,NA49,STAR2005,Adams:2004gp}. Experimenters, 
therefore, have a choice of observables and have in fact reported data using many of these averaging methods. 
The different methods used by several experiments produce results that cannot be trivially compared from one 
experiment to another. The issue is compounded by the fact that these observables have different dependencies on the 
detection efficiency and acceptance. It is therefore of considerable interest to determine how different are the  numerical values 
produced by these different averaging methods and  correlation observable definitions. One can also find out whether 
values measured by an experiment using one observable can be meaningfully compared to those of  different 
experiments or measurements using other observables.

We first note that if particle production follows perfect Poisson statistics, the correlation functions defined in the 
previous section are, by construction, null. The {\it inclusive}, {\it event-wise} and $\tilde C$ observables are thus 
trivially identical in the Poisson fluctuation limit. However, due to energy, momentum and quantum 
number conservation, particle production in general is a non-poissonian stochastic phenomenon, {\it i.e.} the correlation functions 
are non-zero. Below we investigate how different they can be in practice.

In order to investigate the differences between the {\it inclusive}, {\it event-wise} and $\tilde C$ observables, 
we consider a decomposition of the one- and two-particle densities using fixed 
multiplicity densities. We write
\begin{equation}
\rho _1 (p,\eta ) = \sum\limits_{m = 0}^\infty  {P(m)\rho _{1,m} (p,\eta )} 
\label{Eq17}
\end{equation}
\begin{equation}
\rho _2 (p_1 ,\eta _1 ,p_2 ,\eta _2 ) = \sum\limits_{m_1 ,m_2  = 0}^\infty  {P(m_1 ,m_2 )\rho _{2,m_1 ,m_2
(p_1 ,\eta _1 ,p_2 ,\eta _2 )} }
\label{Eq18}
\end{equation}

where $P(m)$ expresses the probability of having a multiplicity `{\it m}' in the measured $\eta$ and $\varphi$ bin. 
Similarly, $P(m_1,m_2)$ corresponds to the probability of finding $m_1$ and $m_2$ particles simultaneously 
in bins ($\eta_1,\varphi_1$) and ($\eta_2,\varphi_2$),  respectively. 
$\rho _{1,m} (p,\eta )$ and $\rho _{2,m} (p_1 ,\eta _1 ,p_2 ,\eta _2 )$ are the single and pair densities for 
a fixed value of multiplicity $m$ ($m_1,m_2$).
Note that in the above expressions we have included a dependency on $\eta$ only for brevity ($\varphi$ 
dependencies can obviously also be included). By construction, one has
\begin{equation}
\int {\rho _{1,m} (p,\eta )} dpd\eta  = m 
\label{Eq19}
\end{equation}
\begin{equation}
\int {\rho _{2,m_1 ,m_2 } (p_1 ,\eta _1 ,p_2 ,\eta _2 )} dp_1 dp_2 d\eta _1 d\eta _2  = m_1 m_2 
\label{Eq20}
\end{equation}

Integration restricted to particle momenta yields the pseudorapidity densities: 
\begin{eqnarray}
\int {\rho _{1,m} (p,\eta )} dp &=& \rho _{1,m} (\eta ) \\ 
\label{Eq21}
\int {\rho _{2,m_1 ,m_2 } (p_1 ,\eta _1 ,p_2 ,\eta _2 )} dp_1 dp_2  &=& \rho _{2,m_1 ,m_2 } (\eta _1 ,\eta _2 ) 
\label{Eq22}
\end{eqnarray}

The averages of the transverse momentum, $p$, and $\Delta p_1 \Delta p_2 $ at fixed multiplicity `{\it m}' are
\begin{eqnarray}
\rho _{_{1,m} }^p (\eta ) &=& \frac{{\int {\rho _{1,m} (p,\eta )pdp} }}{{\rho _{1,m} (\eta )}}\\ 
\label{Eq23}
\rho _{_{2,m_1 ,m_2 } }^{\Delta p\Delta p} (\eta _1 ,\eta _2 ) &=& \frac{{\int {\rho _{2,m_1 ,m_2 } (p_1 ,\eta _1 
,p_2 ,\eta _2 )\Delta p_1 \Delta p_2 dp_1 dp_2 } }}{{\rho _{2,m_1 ,m_2 } (\eta _1 ,\eta _2 )}} 
\label{Eq24}
\end{eqnarray}

Given the multiplicity $m$ ($m_1$, $m_2$) is (are) fixed, the above expressions are valid and identical for the 
{\it inclusive} and {\it event-wise} averaging methods. 
Thus for fixed multiplicity densities the {\it inclusive} and {\it event-wise} averages are truely comparable for analyzing 
actual densities such as occur in $p+p$ and $A+A$ collisions.
Based on Eqs. \ref{Eq12}, \ref{Eq13}, \ref{Eq17} \& \ref{Eq18}, the {\it inclusive} $p_t$ and $\Delta p_t \Delta p_t$ means are 
expressed as follows in terms of the fixed multiplicity densities:
\begin{eqnarray}
\rho _1^p (\eta ) &=& \frac{{\sum\limits_{m = 0}^\infty  {P(m)\rho _{1,m} (\eta )\rho _{_{1,m} }^p (\eta )} 
}}{{\sum\limits_{m = 0}^\infty  {P(m)\rho _{1,m} (\eta )} }} \\ 
\rho _2^{\Delta p_{1}\Delta p_{2}} (\eta _1 ,\eta _2 ) &=& \frac{{\sum\limits_{m_1 ,m_2  = 0}^\infty  {P(m_1 ,m_2 )\int 
{\rho _{2,m_1 ,m_2 } (p_1 ,\eta _1 ,p_2 ,\eta _2 )\Delta p_1 \Delta p_2 dp_1 dp_2 } } }}{{\sum\limits_{m_1 ,m_2  = 
0}^\infty  {P(m_1 ,m_2 )\rho _{2,m} (\eta _1 ,\eta _2 )} }} \nonumber \\ 
  &=& \frac{{\sum\limits_{m_1 ,m_2  = 0}^\infty  {P(m_1 ,m_2 )\rho _{2,m_1 ,m_2 } (\eta _1 ,\eta _2 )\rho 
_{_{2,m_1 ,m_2 } }^{\Delta p\Delta p} (\eta _1 ,\eta _2 )} }}{{\sum\limits_{m_1 ,m_2  = 0}^\infty  {P(m_1 
,m_2 )\rho _{2,m_1 ,m_2 } (\eta _1 ,\eta _2 )} }} 
\label{Eq25}
\end{eqnarray}

The mean $p_t$ and the $\Delta p_t \Delta p_t$ fluctuations within a single wide range of pseudorapidity are given by
\begin{eqnarray}
\left\langle p \right\rangle  &=& \frac{{\sum\limits_{m = 0}^\infty  {P(m)m\rho _{_{1,m} }^p (\eta )} 
}}{{\sum\limits_{m = 0}^\infty  {P(m)\rho _{1,m} (\eta )} }} \\ 
\label{Eq26}
\left\langle {\Delta p\Delta p} \right\rangle  &=& \frac{{\sum\limits_{m = 0}^\infty  {P(m)m(m - 1)\rho _{_{2,m} 
}^{\Delta p\Delta p} (\eta _1 ,\eta _2 )} }}{{\sum\limits_{m = 0}^\infty  {P(m)m} }} 
\label{Eq27}
\end{eqnarray}

The {\it event-wise} mean $p_t$ and the $\Delta p_t \Delta p_t$  are calculated in a similar fashion.  
\begin{eqnarray}
\tilde \rho _1^p (\eta ) &=& \left\langle {\frac{1}{m}\sum\limits_{i = 1}^m {p_i } (\eta )} \right\rangle
= \sum\limits_{m = 0}^\infty  {P(m)\rho _{1,m}^p (\eta )}  \\ 
\label{Eq28}
\tilde \rho _2^{\Delta p\Delta p} (\eta _1 ,\eta _2 ) &=& \left\langle {\frac{1}{{m_1 m_2 }}\sum\limits_{i = 
1}^{m_1 } {\sum\limits_{j \ne i = 1}^{m_2 } {\Delta p_i (\eta _1 )\Delta p_j (\eta _2 )} } } \right\rangle \nonumber \\
&=& \sum\limits_{m_1,m_2  = 0}^\infty  {P(m_1 ,m_2 )\rho _{2,m_1 ,m_2 }^{\Delta p\Delta p} (\eta _1 ,\eta _2 )}  
\label{Eq29}
\end{eqnarray}

For a wide bin of integrated pseudorapidity this becomes				
\begin{eqnarray}
\bar p &=& \left\langle {\frac{1}{m}\sum\limits_{i = 1}^m {p_i } } \right\rangle 
= \sum\limits_{m = 0}^\infty  {P(m)\left\langle p \right\rangle _m }  \\ 
\label{Eq30}
\overline {\Delta p\Delta p}  &=& \left\langle {\frac{1}{{m(m - 1)}}\sum\limits_{i \ne j = 1}^m {p_i p_j } } 
\right\rangle   
= \sum\limits_{m = 0}^\infty  {P(m)\left\langle {\Delta p\Delta p} \right\rangle _m }  
\label{Eq31}
 \end{eqnarray}

Lastly, $\tilde C$ expressed in terms of fixed multiplicity densities is calculated as follows:
\begin{eqnarray}
\lefteqn{C = \frac{{\left\langle {m(m - 1)\left\langle {p_1 p_2 } \right\rangle _m } \right\rangle }}{{\left\langle m 
\right\rangle ^2 }} - \left\langle p \right\rangle ^2} \nonumber \\
&=& \frac{{\sum\limits_{m = 0}^\infty  {P(m)m(m - 1)\left\langle {p_1 p_2 } \right\rangle _m } }}{{\left\langle m \right\rangle ^2 }} - \left\langle p \right\rangle ^2 
\label{Eq32}
\end{eqnarray}

Though the {\it inclusive}, {\it event-wise} and $\tilde C$ quantities are manifestly different, they exhibit interesting similarities. We 
first consider the difference between the {\it inclusive} mean, $\left\langle p \right\rangle$, and the {\it event-wise} mean, $p_{t}$, for 
large integrated ranges of pseudorapidity. Assuming $\left\langle p \right\rangle _m$ 
 is roughly constant over the multiplicity range of interest, one finds
\begin{eqnarray}
&&\left\langle p \right\rangle  = \frac{{\sum\limits_{m = 0}^\infty  {P(m)m\left\langle p \right\rangle _m } 
}}{{\left\langle m \right\rangle }} \approx \left\langle p \right\rangle _m \frac{{\sum\limits_{m = 0}^\infty  
{P(m)m} }}{{\left\langle m \right\rangle }} = \left\langle p \right\rangle _m  \\ 
\label{Eq33}
&&\bar p = \sum\limits_{m = 0}^\infty  {P(m)\left\langle p \right\rangle _m }  \approx \left\langle p \right\rangle 
_m \sum\limits_{m = 0}^\infty  {P(m)}  = \left\langle p \right\rangle _m  
\label{Eq34}
\end{eqnarray}

{\it i.e.} the {\it inclusive} and {\it event-wise} mean $p_{t}$ should be approximately equal. This approximation is most likely valid 
for large domains of integration in pseudorapidity and in central $A+A$ collisions, which yield large multiplicities. 
In this case, it is reasonable to expect that $\left\langle p \right\rangle _m$ varies slowly with {\it m} 
and one thus expects the difference between the two observables to be the smallest. On the other hand for 
small domains of integration or small multiplicity $\left\langle p \right\rangle _m$
is more likely to depend on {\it m} and consequently one expects the two average methods to exhibit larger 
numerical differences.

The {\it inclusive}, {\it event-wise} and $\tilde C$ two-particle correlations can be approximated in the same manner.  One finds 
for integral, or wide bin averages:
\begin{eqnarray}
&&\left\langle {\Delta p_1 \Delta p_2 } \right\rangle  \approx \left\langle {\Delta p_1 \Delta p_2 } \right\rangle _m 
\frac{{\sum\limits_{m = 0}^\infty  {P(m)m(m - 1)} }}{{\left\langle {m(m - 1)} \right\rangle }} = \left\langle {\Delta p_1 \Delta p_2 } \right\rangle _m  \\ 
\label{Eq35}
&& \overline {\Delta p_1 \Delta p_2 }  \approx \left\langle {\Delta p_1 \Delta p_2 } \right\rangle _m \sum\limits_{m 
= 0}^\infty  {P(m)}  = \left\langle {\Delta p_1 \Delta p_2 } \right\rangle _m  \\ 
\label{Eq36}
\tilde C &=& \frac{{\sum\limits_{m = 0}^\infty  {P(m)m(m - 1)\left( {\left\langle {\Delta p_1 \Delta p_2 } \right\rangle 
_m  + \left\langle p \right\rangle _m ^2 } \right)} }}{{\left\langle m \right\rangle ^2 }} - \left\langle p 
\right\rangle ^2  \nonumber \\ 
&=& \left\langle {\Delta p_1 \Delta p_2 } \right\rangle \frac{{\left\langle {m(m - 1)} \right\rangle }}{{\left\langle 
m \right\rangle ^2 }} + \frac{{\sum\limits_{m = 0}^\infty  {P(m)m(m - 1)\left\langle p \right\rangle _m ^2 } 
}}{{\left\langle m \right\rangle ^2 }} - \left\langle p \right\rangle ^2 \nonumber \\ 
&\approx& \left\langle {\Delta p_1 \Delta p_2 } \right\rangle \frac{{\left\langle {m(m - 1)} \right\rangle 
}}{{\left\langle m \right\rangle ^2 }} + \left\langle p \right\rangle ^2 \left( {\frac{{\left\langle {m(m - 1)} 
\right\rangle }}{{\left\langle m \right\rangle ^2 }} - 1} \right) 
\label{Eq37}
\end{eqnarray}

where we have again assumed both $\left\langle p \right\rangle _m$ and $\left\langle {\Delta p_1 \Delta p_2 } \right\rangle _m$ 
vary slowly with `{\it m}'. One finds that in the limit where $\left\langle p \right\rangle _m$
and $\left\langle {\Delta p_1 \Delta p_2 } \right\rangle _m$
are independent of $m$, the {\it inclusive} and {\it event-wise} observables are strictly equal. In practice, the equality is 
only approximate given the average particle momentum is a function of event 
multiplicity. Additionally, given that for large values of $m$, and relatively narrow range of multiplicity values,  
the ratio $\frac{{\left\langle {m(m - 1)} \right\rangle }}{{\left\langle m \right\rangle ^2 }}$
is, typically near unity. Therefore, one concludes that $\tilde C$ is also roughly equal to $\left\langle {\Delta p_1 \Delta p_2 } \right\rangle$
 for large multiplicities. The above arguments and approximate equalities hold for integral correlations (or 
fluctuations) but are weaker for differential correlations. Indeed, in the case of differential correlations, the number 
of particle in a given pseudorapidity bin is quite small. The approximations used in Eqs. \ref{Eq19} \& \ref{Eq20} are thus likely to be 
inappropriate and one should expect significant differences between the three observables; 
particularly between the {\it inclusive} average and $\tilde C$. 
Any comparison of results from experiments reported on the basis of different observables is thus at best qualitative. We explore 
these differences further in Sect. \ref{Sect:Model} on the basis of PYTHIA simulations.

\section{Model Predictions}
\label{Sect:Model}

\begin{figure}[!htp]
\centering
\resizebox{8.7cm}{6.5cm}{\includegraphics{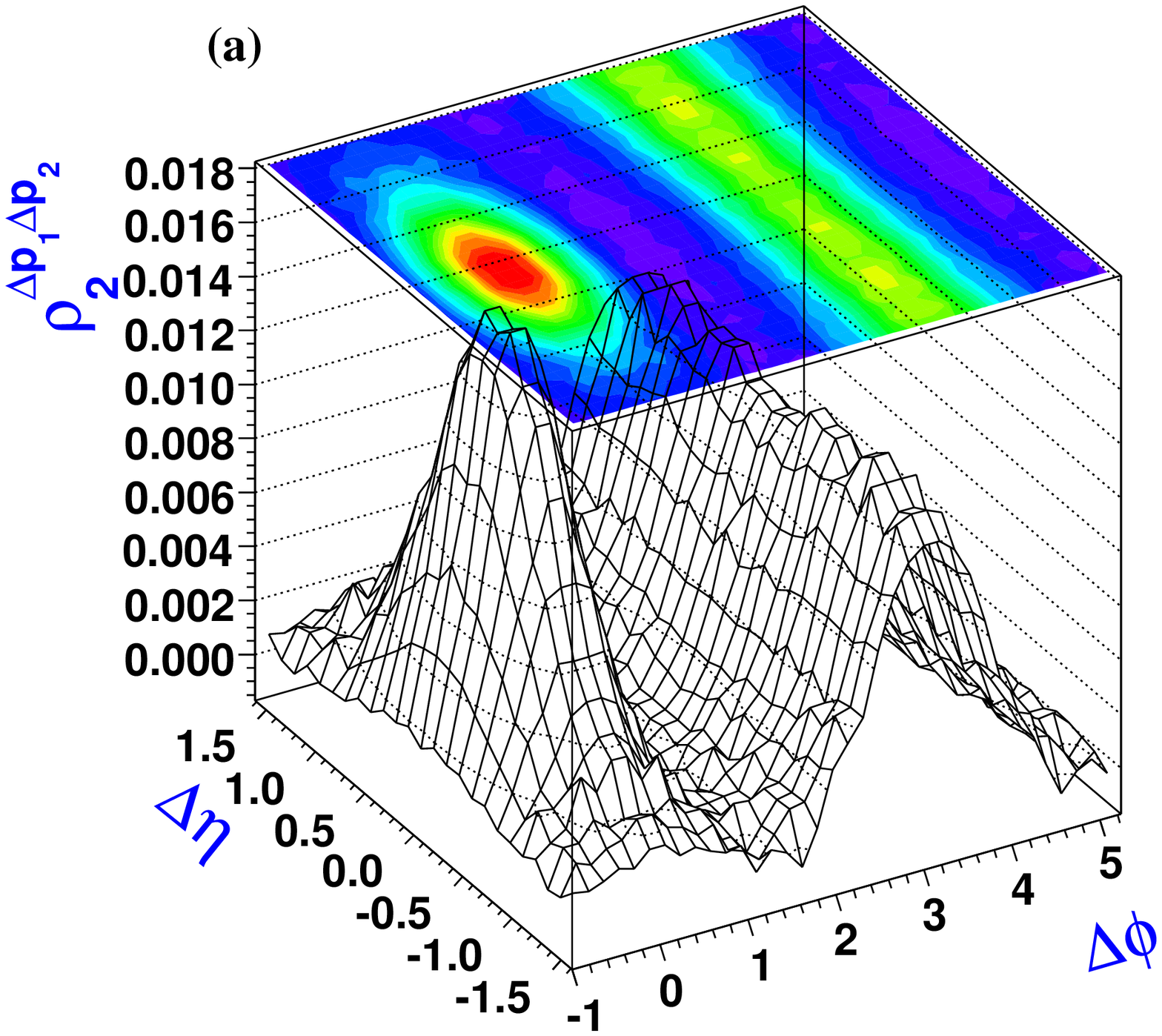}}
\resizebox{8.7cm}{6.5cm}{\includegraphics{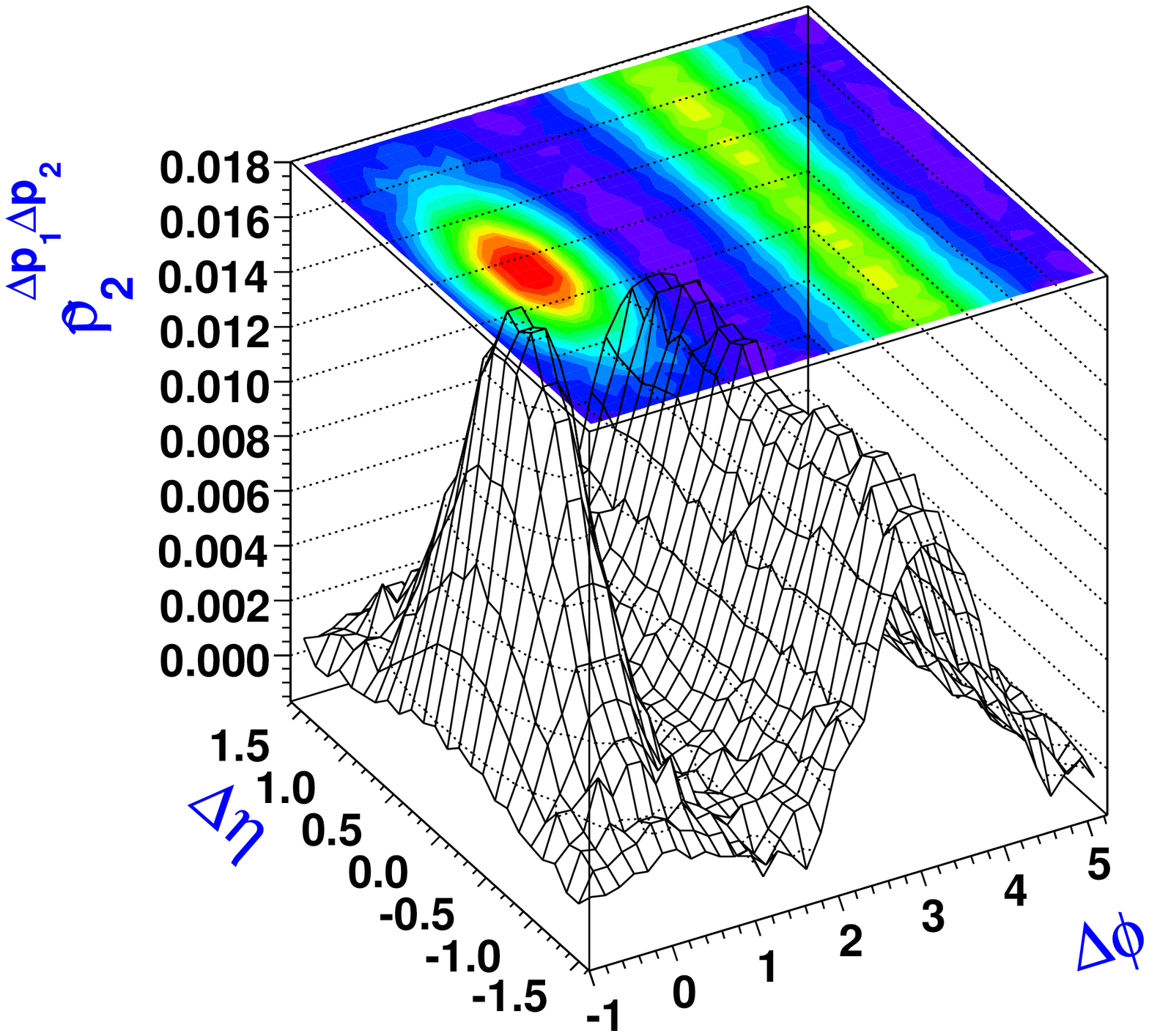}}
\resizebox{8.7cm}{6.5cm}{\includegraphics{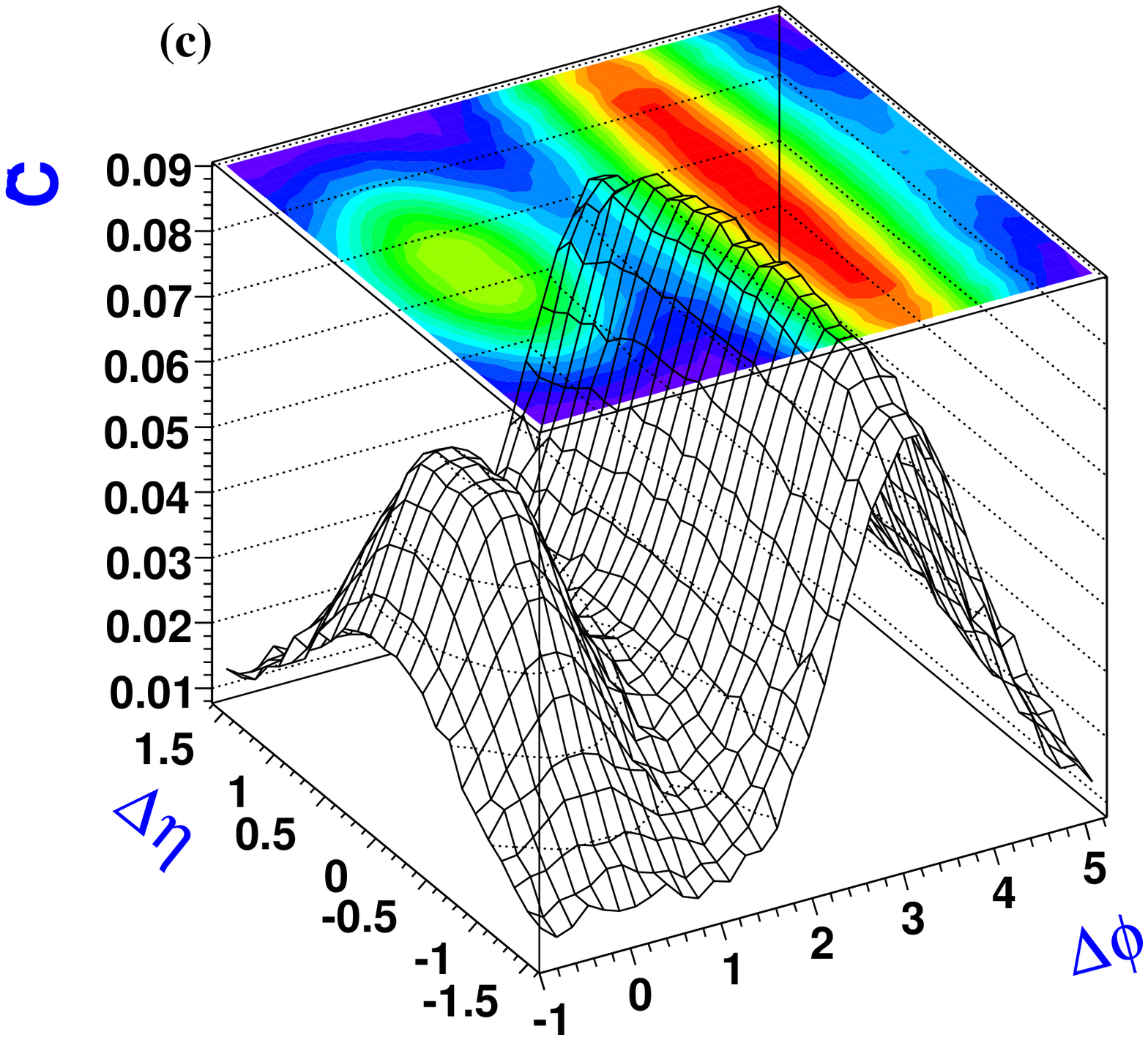}}
\caption[]{(Color online) Comparison of (a) {\it inclusive}, $\rho _{_2 }^{\Delta p_1 \Delta p_2 } \left( {\Delta \eta ,\Delta \varphi } \right)$, 
(b) {\it event-wise}, $\tilde \rho _{_2 }^{\Delta p_1 \Delta p_2 } \left( {\Delta \eta ,\Delta \varphi } \right)$ and (c) $\tilde C$ transverse 
momentum correlation functions obtained with $p+p$ collisions at $\sqrt{s}=200$ GeV generated with PYTHIA. }
\label{fig:1}
\end{figure}

We use PYTHIA (version 6.22) \cite{Pythia} to simulate $p+p$ collisions at $\sqrt{s}=200$ GeV and carry out 
calculations of transverse momentum correlations based on the observables introduced in Sect 
\ref{Sect:PtCorrelations}. Five million minimum bias PYTHIA events were integrated to produce plots shown in the 
following. Figures \ref{fig:1}(a-c) show the {\it inclusive}, $\rho _{_2 }^{\Delta p_1                                        
\Delta p_2 } \left( {\Delta \eta ,\Delta \varphi } \right)$, {\it event-wise}, $\tilde \rho _{_2 }^{\Delta p_1 \Delta p_2 } \left( 
{\Delta \eta ,\Delta \varphi } \right)$ and $\tilde C$ correlation 
functions, respectively, for particles in the pseudorapidity range $|\eta|<1$ and transverse momentum range $0.2<p_t<2.0$ GeV/c. 
The two-particle correlations are plotted as a function of the particles' relative pseudorapidity, $\Delta\eta$, and 
azimuthal angles, $\Delta\varphi$, using 31 and 36 bins, respectively. 

The three correlation functions show qualitative similarities: all three observables exhibit a peak structure near 
$\Delta\eta~=~\Delta\varphi~=~0$ (near side) and a ridge-like structure at $\Delta\varphi=\pi$ (away side) which extends over the full 
$\Delta\eta$ range. These shapes are qualitatively similar to that of density correlations predicted by 
PYTHIA \cite{Pruneau07}. While the {\it inclusive} and {\it event-wise} correlation functions are identical to each other with 
a difference of only 0.2\%  
(statistical errors are of the order of $\pm$0.001), the observable $\tilde C$ exhibits characteristically different strength and shape. 
In contrast to {\it inclusive} and {\it event-wise} observables, the away side 
ridge-like structure has larger magnitude than the near side. One also finds that $\tilde C$ has a strength 
five times larger than the other two observables. A detailed study of $\tilde C$ reveals that the shape is largely
determined by density correlations, $\frac{\rho_{2}}{\rho_1,\rho_1}$. In contrast to {\it inclusive} and {\it event-wise} measures, which
are designed to minimize the number density contribution $\tilde C$ is sensitive to the variation of the $p_{t}$ 
of the particles as well as the number density. As a result we conclude that the differences between the three observables 
stem from their specific definitions and hence they lead to very different numerical results. 

\begin{figure}[!htp]
\centering
\resizebox{8.7cm}{6.5cm}{\includegraphics{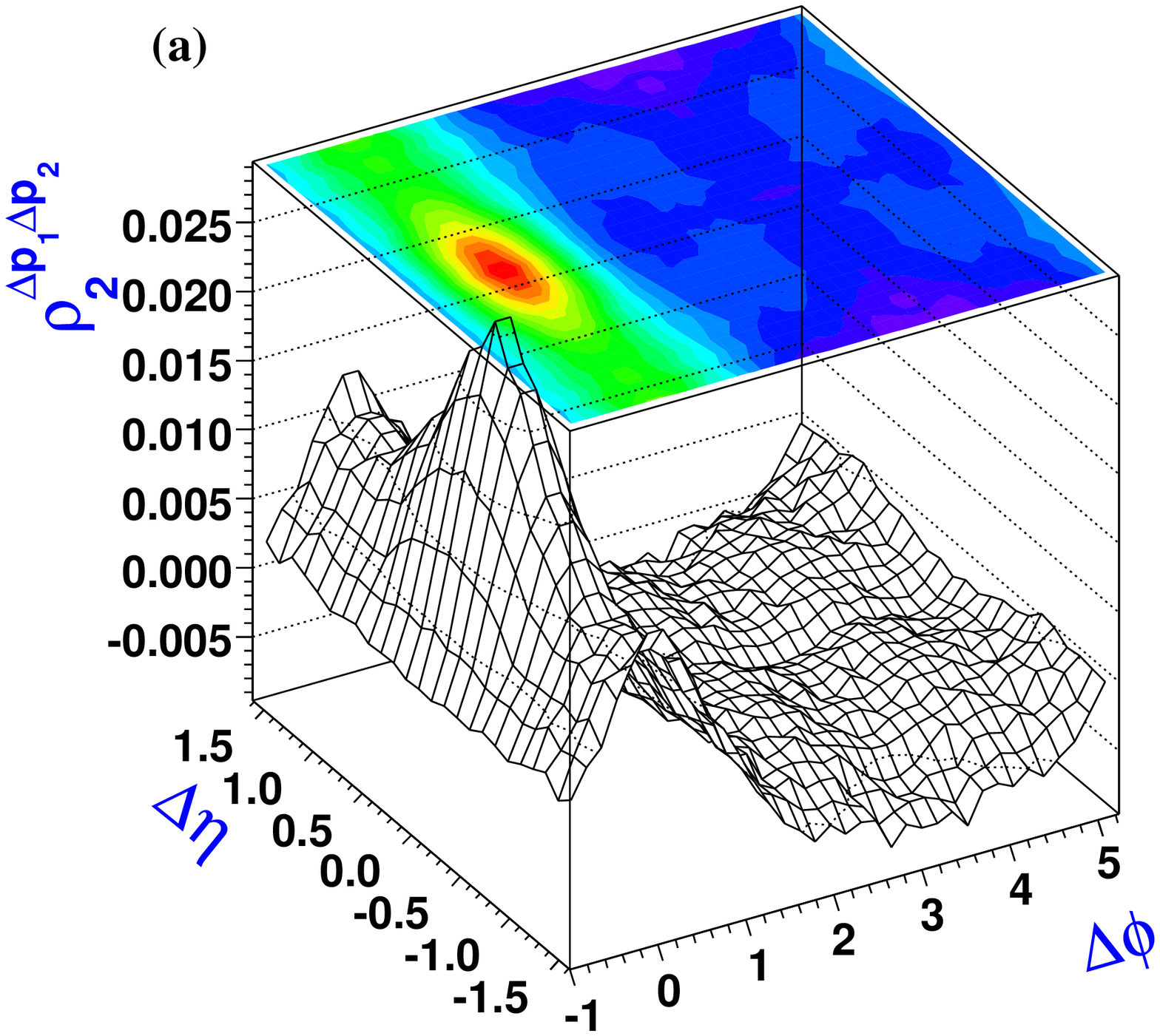}}
\resizebox{8.7cm}{6.5cm}{\includegraphics{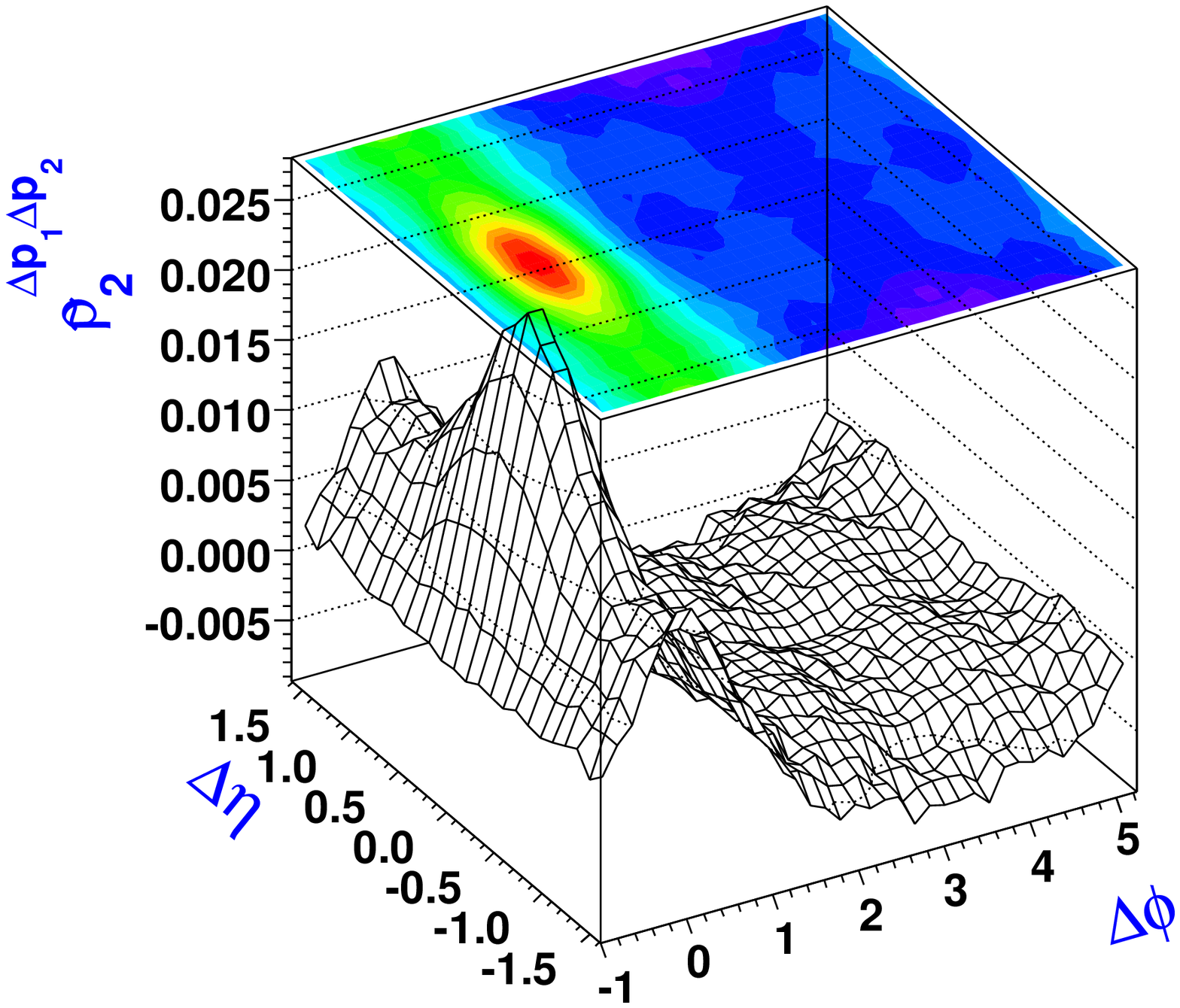}}
\resizebox{8.7cm}{6.5cm}{\includegraphics{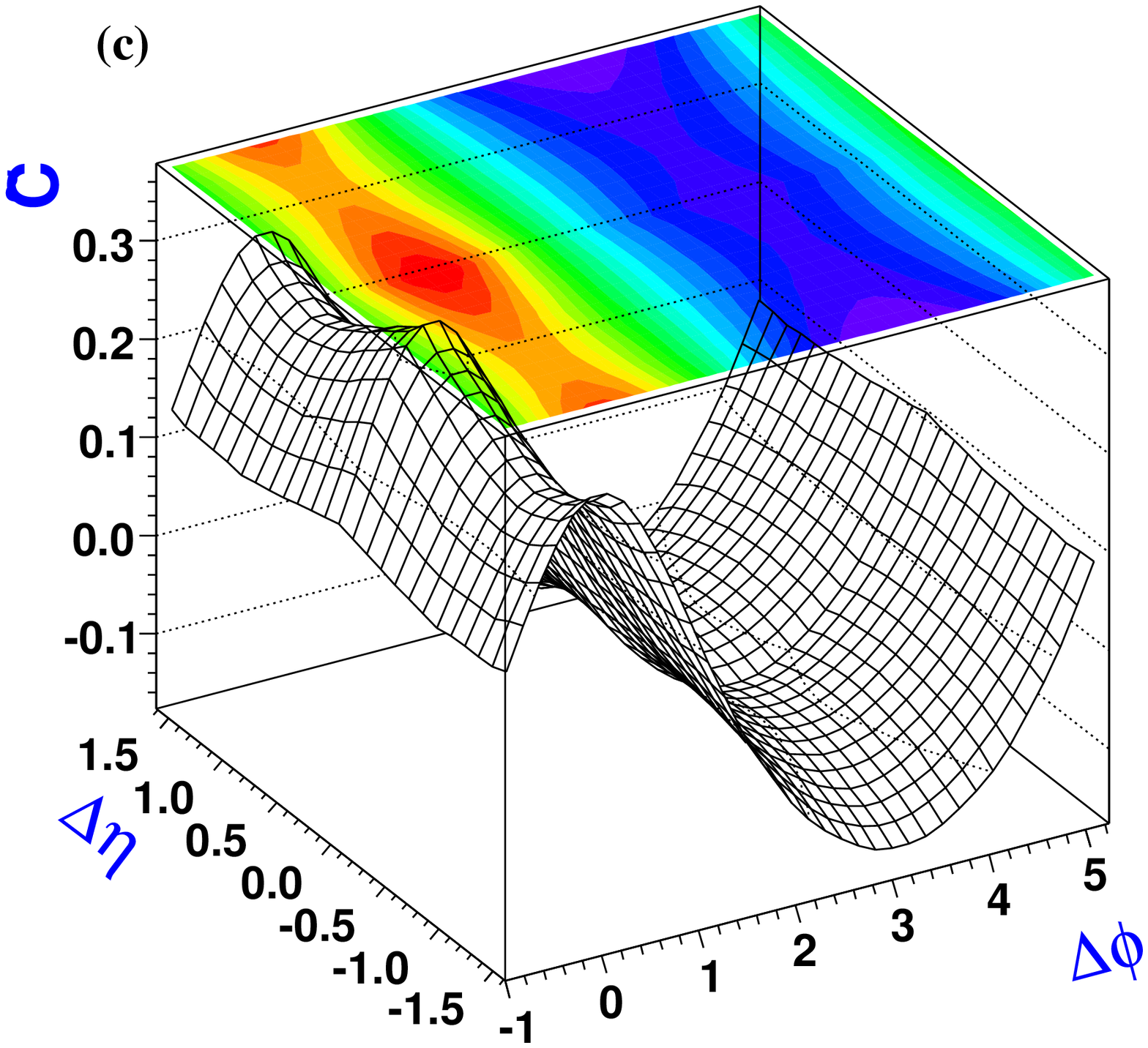}}
\caption[]{(Color online) Comparison of (a) {\it inclusive}, $\rho _{_2 }^{\Delta p_1 \Delta p_2 } \left( {\Delta \eta ,\Delta \varphi } \right)$, 
(b) {\it event-wise}, $\tilde \rho _{_2 }^{\Delta p_1 \Delta p_2 } \left( {\Delta \eta ,\Delta \varphi } \right)$ 
and (c) $\tilde C$ transverse momentum correlation functions obtained with radially boosted ($\beta = $ 0.3) $p+p$ collisions at 
$\sqrt{s}=200$ GeV.}
\label{fig:2}
\end{figure}

Measurements of two-particle correlation functions, $\rho _2 (\Delta \eta ,\Delta \varphi )$, 
in $Au + Au$ collisions reveal the presence of a strong ridge-like structure on the near-side, $\Delta \varphi  
\approx 0$, and small correlation yield on the away-side, $\Delta \varphi  \approx \pi $,
while elliptic flow appears as a rather modest contribution to the overall yield of the correlation function. 
Authors of Ref. \cite{Pruneau07} have suggested the change in the shape of the correlation 
functions, in which a ridge-like structure shifts from $\Delta \varphi  \approx \pi $ to 
$\Delta \varphi  \approx 0$, might in part be caused by strong radial flow present in mid to central $A+A$ collisions. 
It is thus interesting to test the radial flow scenario by considering the effects of radial flow on transverse momentum correlations.
 Figures \ref{fig:2}(a-c) show the {\it inclusive}, {\it event-wise} and $\tilde C$ correlation functions 
calculated for radially boosted $p+p$ events. All particles produced by 
a given $p+p$ event are boosted radially in the transverse plane with factor $\beta=0.3$. Radial flow imparts kinematic focussing, as a result 
of which all the particles from each $p+p$ collisions are pushed in the same direction and hence become correlated in azimuth. 
Indeed one finds a significant correlated yield at large pair separation in pseudorapidity and narrow separation in azimuth 
on the near side of all the three observables. The {\it inclusive} and {\it event-wise} measures are qualitatively similar with a peak and 
ridge-like structure on the near side and diminished correlations on the away side. However, $\tilde C$ differs quantitatively. The 
away side correlations are reduced significantly and the near side peak-like structure is more prominent 
vis-a-vis {\it inclusive} and {\it event-wise} observables. 
Also, the strength of $\tilde C$ for boosted $p+p$ events is nearly 10 times larger than the strength of the other two observables. 
These simulations show that radial flow in $A + A$ collisions can produce a near-side ridge extending in pseudorapidity in transverse momentum 
correlations as well as in two-particle density correlations.

\section{Scaling with number of participants in heavy ion collisions}
\label{Sect:Scaling}

In this section, we consider the $\Delta p_1 \Delta p_2$ dependence on the number of participants in heavy ion 
collisions. We use as a reference collisions where nucleon-nucleon interactions are independent of one 
another and we neglect re-scattering of the produced particles. 

Under such assumptions, the two-particle density in $A+A$ collisions at fixed number of participants, $N$, may be written 
as follows:
\begin{equation}
\rho _{2,AA}  = N(N - 1)\rho _{1,pp} \rho _{1,pp}  + N\rho _{2,pp} 
\label{Eq38}
\end{equation}

The {\it inclusive} differential transverse momentum correlation function at fixed $N$ in $A+A$ collisions, 
$\left. {\rho _{2,AA}^{\Delta p_1 \Delta p_2 } (\eta _1 ,\eta _2 )} \right|_N $ is then:
\begin{eqnarray}
\left. {\rho _{_{2,AA} }^{\Delta p_1 \Delta p_2 } (\eta _1 ,\eta _2 )} \right|_N  &=& \frac{{N(N - 1)\rho _{1,pp} 
(\eta _1 )\rho _{1,pp} (\eta _2 )\rho _{_{1,pp} }^{\Delta p_t } (\eta _1 )\rho _{_{1,pp} }^{\Delta p_t } (\eta _2 ) 
+ N\rho _{2,pp} (\eta _1 ,\eta _2 )\rho _{_{2,pp} }^{\Delta p_t \Delta p_t } (\eta _1 ,\eta _2 )}}{{N(N - 1)\rho 
_{1,pp} (\eta _1 )\rho _{1,pp} (\eta _2 ) + N\rho _{2,pp} (\eta _1 ,\eta _2 )}} \nonumber \\
&=& \frac{{N\rho _{2,pp} (\eta _1 ,\eta _2 )\rho _{_{2,pp} }^{\Delta p_t \Delta p_t } (\eta _1 ,\eta _2 )}}{{N(N - 
1)\rho _{1,pp} (\eta _1 )\rho _{1,pp} (\eta _2 ) + N\rho _{2,pp} (\eta _1 ,\eta _2 )}} 
\label{Eq39}
\end{eqnarray}

where we have included, for simplicity, only dependencies on the pseudorapidity of the particles. 
The simplification obtained on the second line in Eq. \ref{Eq39} arises because the average of $\Delta p_t$ (first term of the 
numerator) is by construction null.  Experimentally, one cannot constrain the collision impact parameter, so one 
must account for fluctuations in the number of participants. The {\it inclusive} differential correlation function is 
then
\begin{eqnarray}
\rho _{_{2,AA} }^{\Delta p_1 \Delta p_2 } (\eta _1 ,\eta _2 ) &=& \frac{{\left\langle N \right\rangle \rho _{2,pp} 
(\eta _1 ,\eta _2 )\rho _{_{2,pp} }^{\Delta p_t \Delta p_t } (\eta _1 ,\eta _2 )}}{{\left\langle {N(N - 1)} 
\right\rangle \rho _{1,pp} (\eta _1 )\rho _{1,pp} (\eta _2 ) + \left\langle N \right\rangle \rho _{2,pp} (\eta _1 ,\eta 
_2 )}} \nonumber \\ 
&=& \frac{{\left\langle N \right\rangle }}{{\left\langle {N(N - 1)} \right\rangle {{\rho _{1,pp} (\eta _1 )\rho 
_{1,pp} (\eta _2 )} \mathord{\left/
 {\vphantom {{\rho _{1,pp} (\eta _1 )\rho _{1,pp} (\eta _2 )} {\rho _{2,pp} (\eta _1 ,\eta _2 )}}} \right.
 \kern-\nulldelimiterspace} {\rho _{2,pp} (\eta _1 ,\eta _2 )}} + \left\langle N \right\rangle }}\rho _{_{2,pp} 
}^{\Delta p_t \Delta p_t } (\eta _1 ,\eta _2 ) 
\label{Eq40}
\end{eqnarray}

Fluctuations in the number of $p+p$ interactions have Poisson statistics in the context of our independent 
collisions model. It implies:
$\left\langle {N(N - 1)} \right\rangle  = \left\langle N \right\rangle ^2  \gg \left\langle N \right\rangle$
Additionally, since
${{\rho _{_2 }^{(pp)} (\eta _1 ,\eta _2 )} \mathord{\left/
 {\vphantom {{\rho _{_2 }^{(pp)} (\eta _1 ,\eta _2 )} {\rho _{_1 }^{(p)} (\eta _1 )\rho _{_1 }^{(p)} (\eta _2 
)}}} \right.
 \kern-\nulldelimiterspace} {\rho _{_1 }^{(p)} (\eta _1 )\rho _{_1 }^{(p)} (\eta _2 )}} \approx 1$
one expects
\begin{equation}
\rho _{2,AA}^{\Delta p_1 \Delta p_2 } (\eta _1 ,\eta _2 ) \approx \frac{1}{{\left\langle N \right\rangle }}\rho 
_{2,pp}^{\Delta p_t \Delta p_t } (\eta _1 ,\eta _2 )
\label{Eq41}
\end{equation}

{\it i.e.} the {\it inclusive} transverse momentum correlations in $A+A$ collisions should be approximately proportional 
to those measured in $p+p$ collisions and inversely proportional to the number of participants.

\begin{figure}[!htp]
\centering
\resizebox{8.7cm}{6.5cm}{\includegraphics{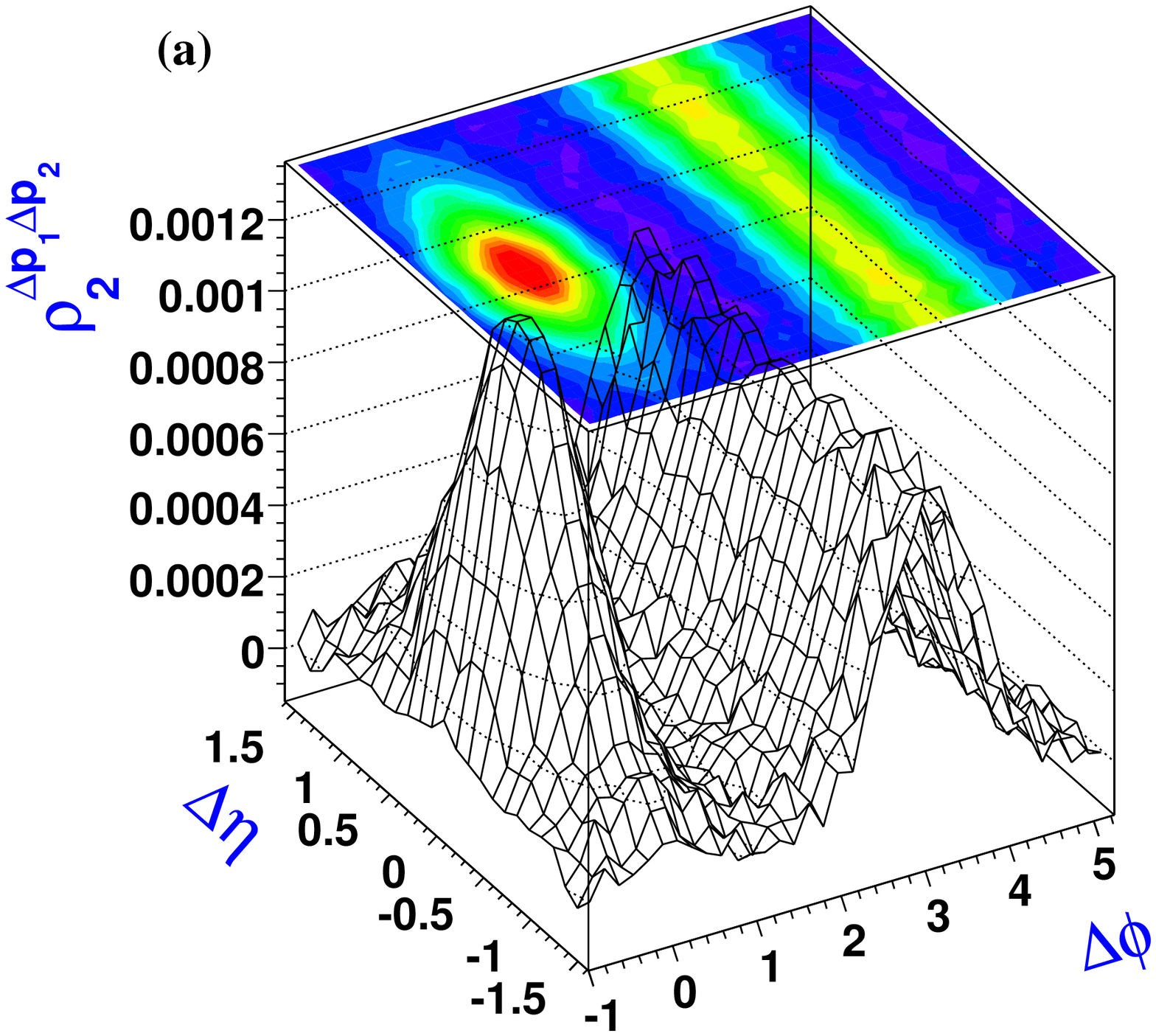}}
\resizebox{8.7cm}{6.5cm}{\includegraphics{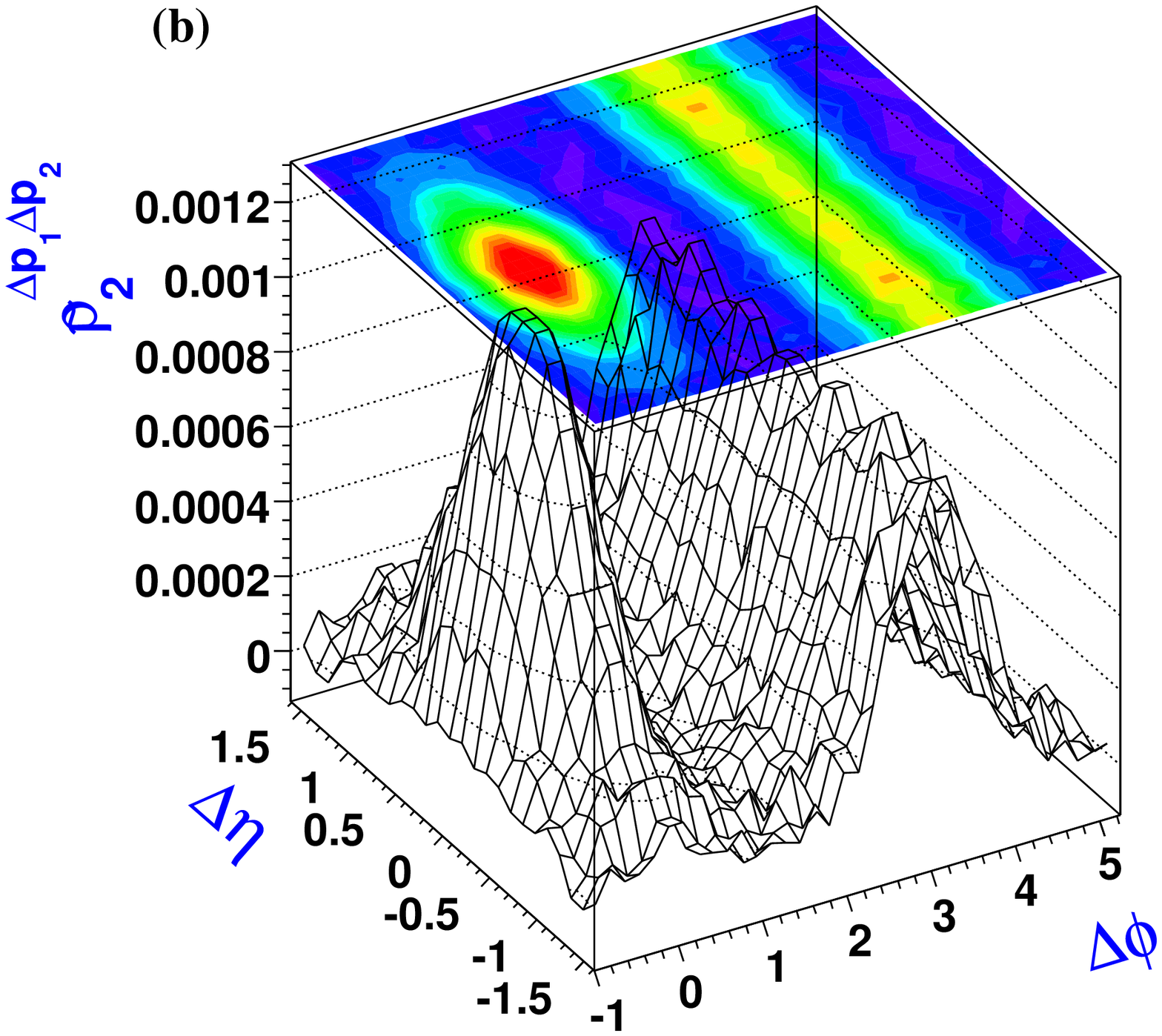}}
\resizebox{8.7cm}{6.5cm}{\includegraphics{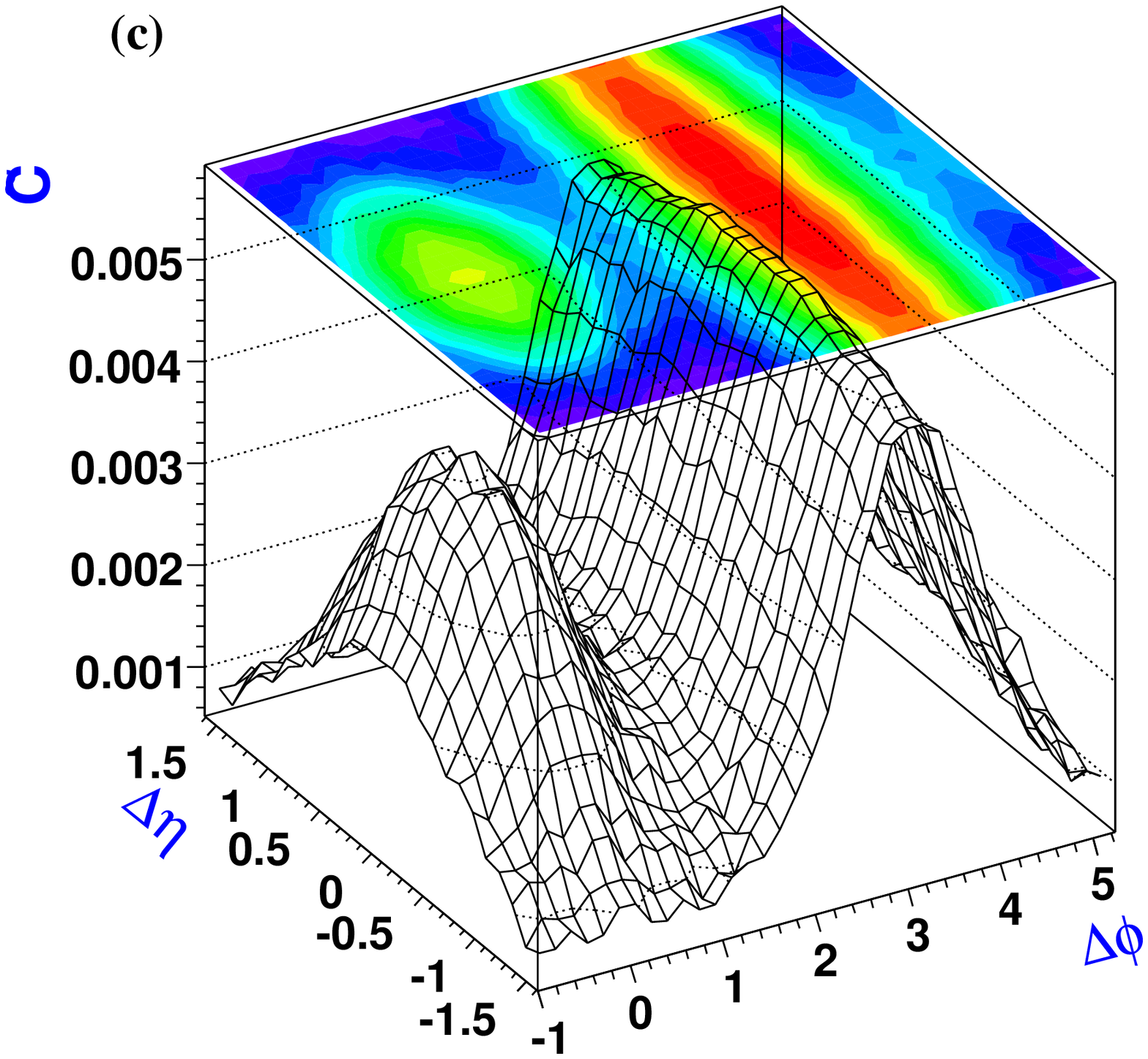}}
\caption[]{(Color online) Comparison of (a) {\it inclusive}, $\rho _{_2 }^{\Delta p_1 \Delta p_2 } \left( {\Delta \eta ,\Delta \varphi } \right)$, (b)
 {\it event-wise}, $\tilde \rho _{_2 }^{\Delta p_1 \Delta p_2 } \left( {\Delta \eta ,\Delta \varphi } \right)$ and (c) $\tilde C$ transverse momentum 
correlation functions obtained with superposition of {\it N} = 15 independent $p+p$ collisions at $\sqrt{s}=200$ GeV generated with 
PYTHIA.}
\label{fig:3}
\end{figure}

We check this result by artificially constructing $A+A$ events on the basis of a superposition of independent {\it N}~=~15
 $p+p$ collisions generated with PYTHIA. Figures \ref{fig:3}(a-c) display correlation functions
obtained with {\it N}~=~15 independent $p+p$ collisions modeling $A+A$ collisions with 30 participants. A total of two million 
minimum bias events were integrated to produce these results.

The most prominent feature observed for all the observables is the dilution of signal strength. The signal strength is diluted by a 
factor of 15 as compared to that found in $p+p$ collisions (shown in Figure \ref{fig:1}). This effect is attributed to 15 times larger 
multiplicity produced for {\it N}~=~15 independent $p+p$ collisions. The shape of {\it inclusive} and {\it event-wise} observables, 
with a peak-like structure 
on the near side and a ridge-like structure on the away side is still observed, as previously observed for Figure \ref{fig:1}(a,b). Similarly 
we observe the same shape of $\tilde C$, {\it i.e.} larger amplitude of the away side ridge vis-a-vis the near side peak.  

\begin{figure}[!htp]
\centering
\resizebox{8.7cm}{6.5cm}{\includegraphics{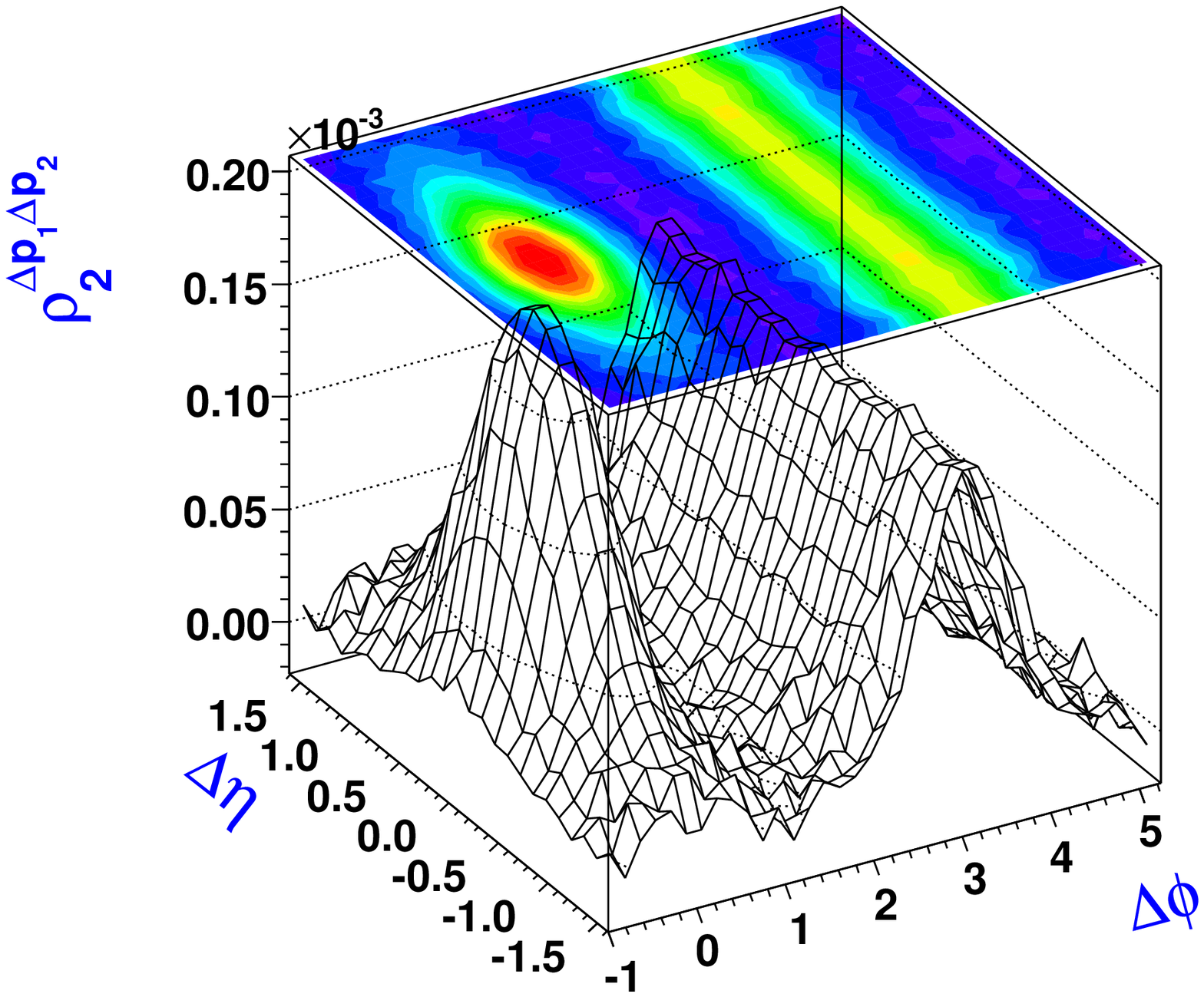}}
\resizebox{8.7cm}{6.5cm}{\includegraphics{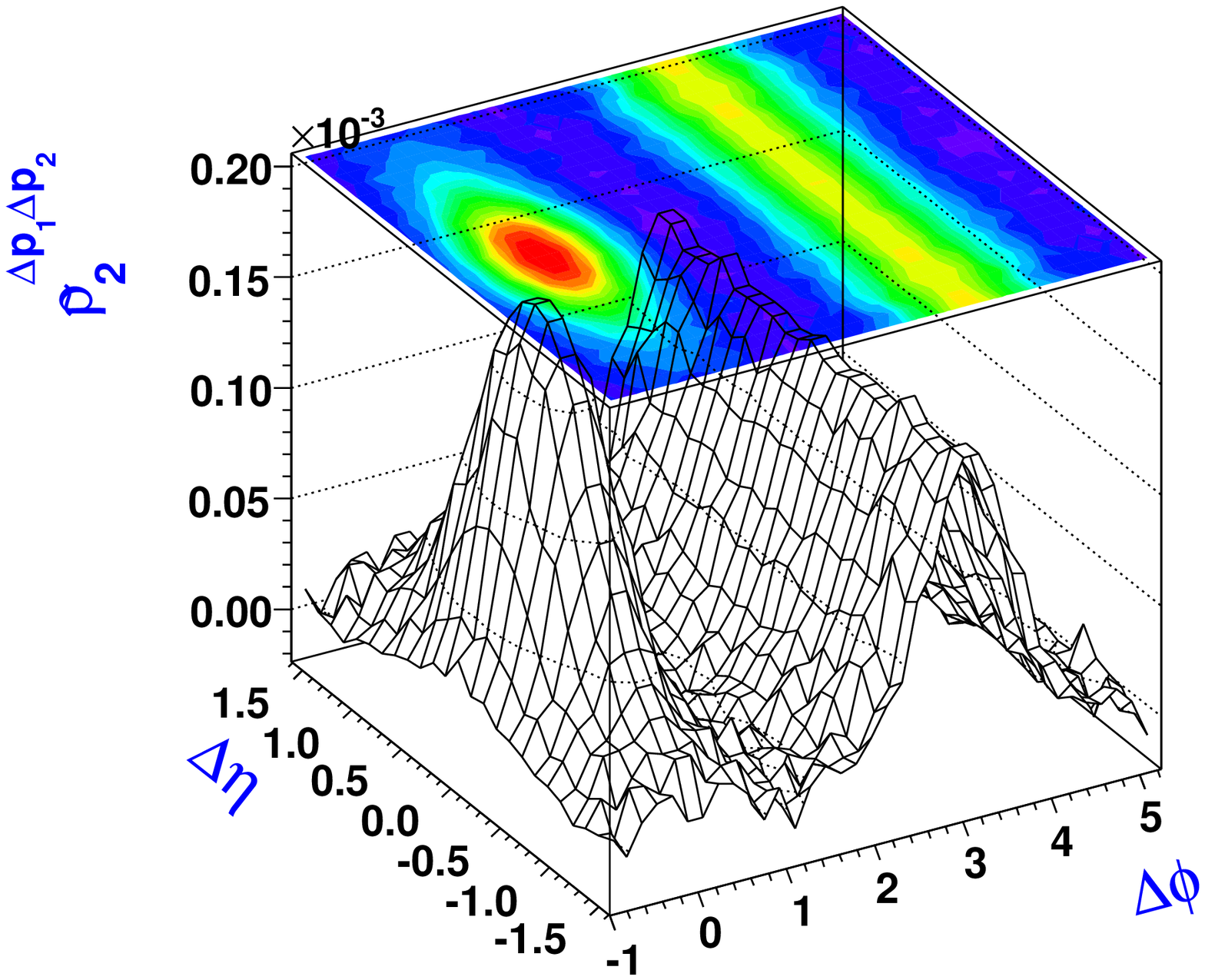}}
\resizebox{8.7cm}{6.5cm}{\includegraphics{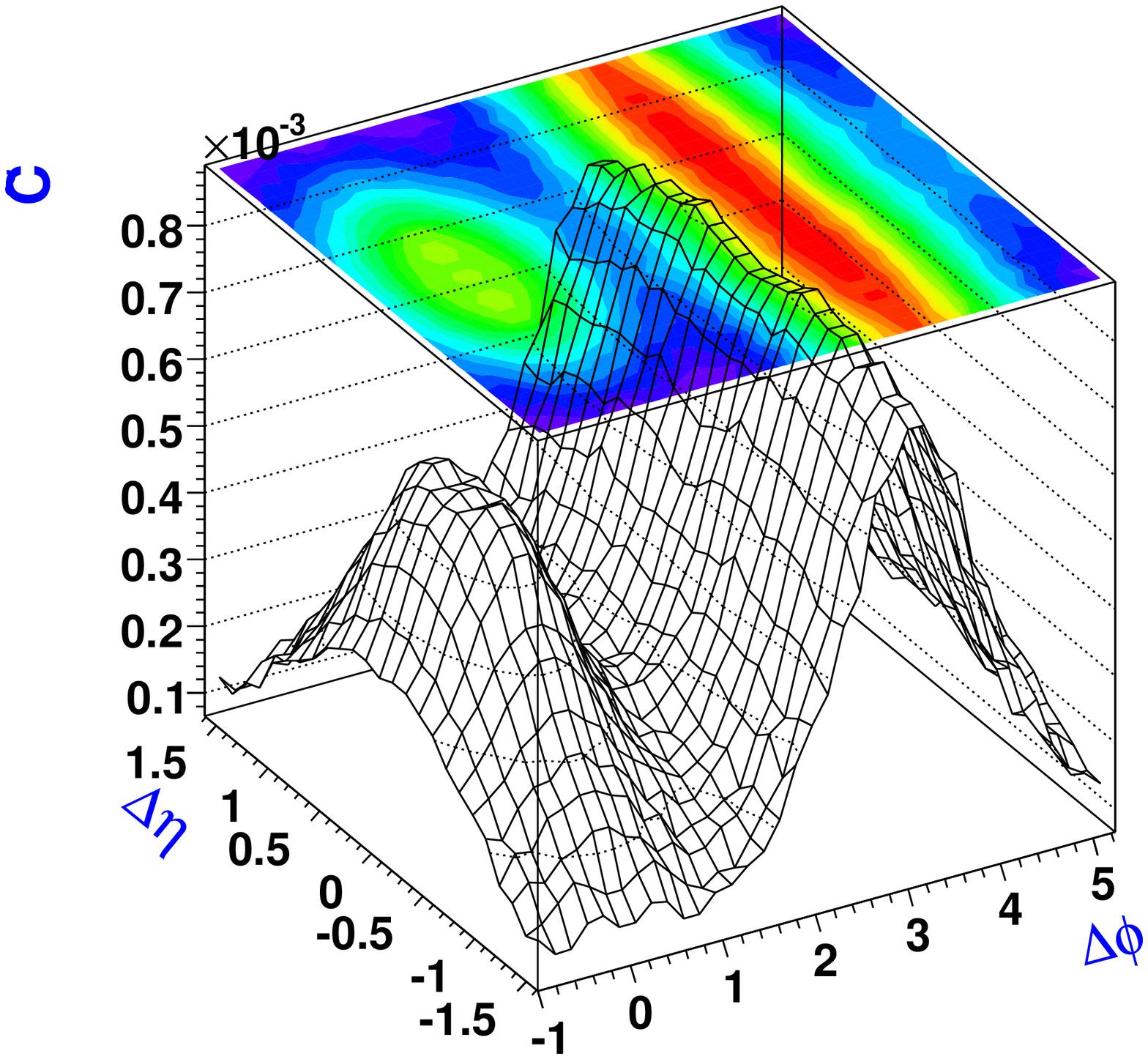}}
\caption[]{(Color online) Comparison of (a) {\it inclusive}, $\rho _{_2 }^{\Delta p_1 \Delta p_2 } \left( 
{\Delta \eta ,\Delta \varphi } \right)$, (b) event-wise, $\tilde \rho _{_2 }^{\Delta p_1 \Delta p_2 } 
\left( {\Delta \eta ,\Delta \varphi } \right)$ and (c) $\tilde C$ transverse momentum correlation functions 
obtained with radially boosted ($\beta = $ 0.3) {\it N} = 100 $p+p$ collisions at $\sqrt{s}=200$ GeV, 
generated with PYTHIA. }
\label{fig:44}
\end{figure}

We also present results on $A+A$ collisions produced by the superposition of {\it N}~=~100 $p+p$ collisions, i.e., 200 
participants. Two million minimum bias events were integrated to present the results in Figure \ref{fig:44}(a-c). As expected, we observe the 
signal strength is diluted by a factor of 100. 

\begin{figure}[!htp]
\centering
\resizebox{8.7cm}{6.5cm}{\includegraphics{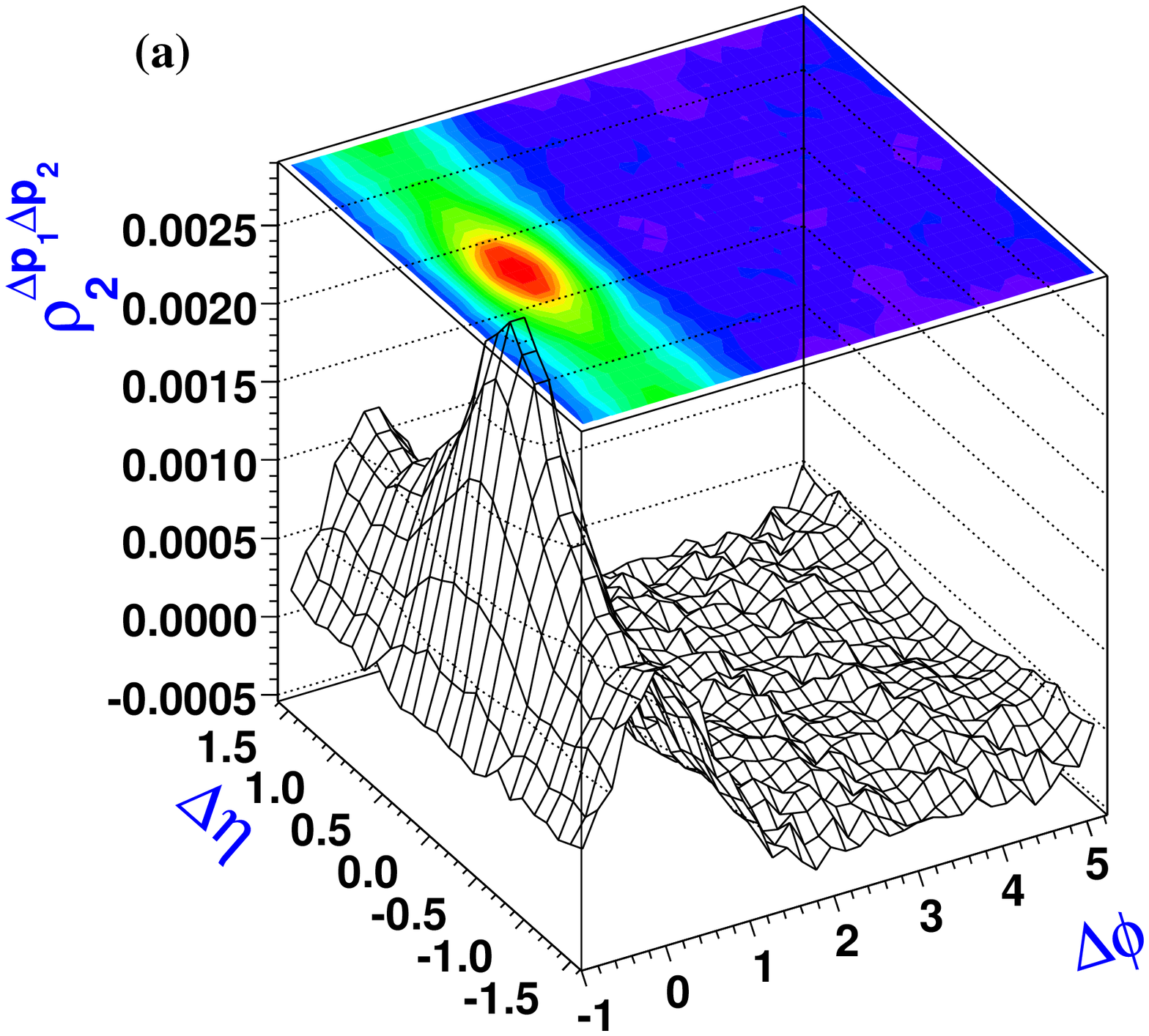}}
\resizebox{8.7cm}{6.5cm}{\includegraphics{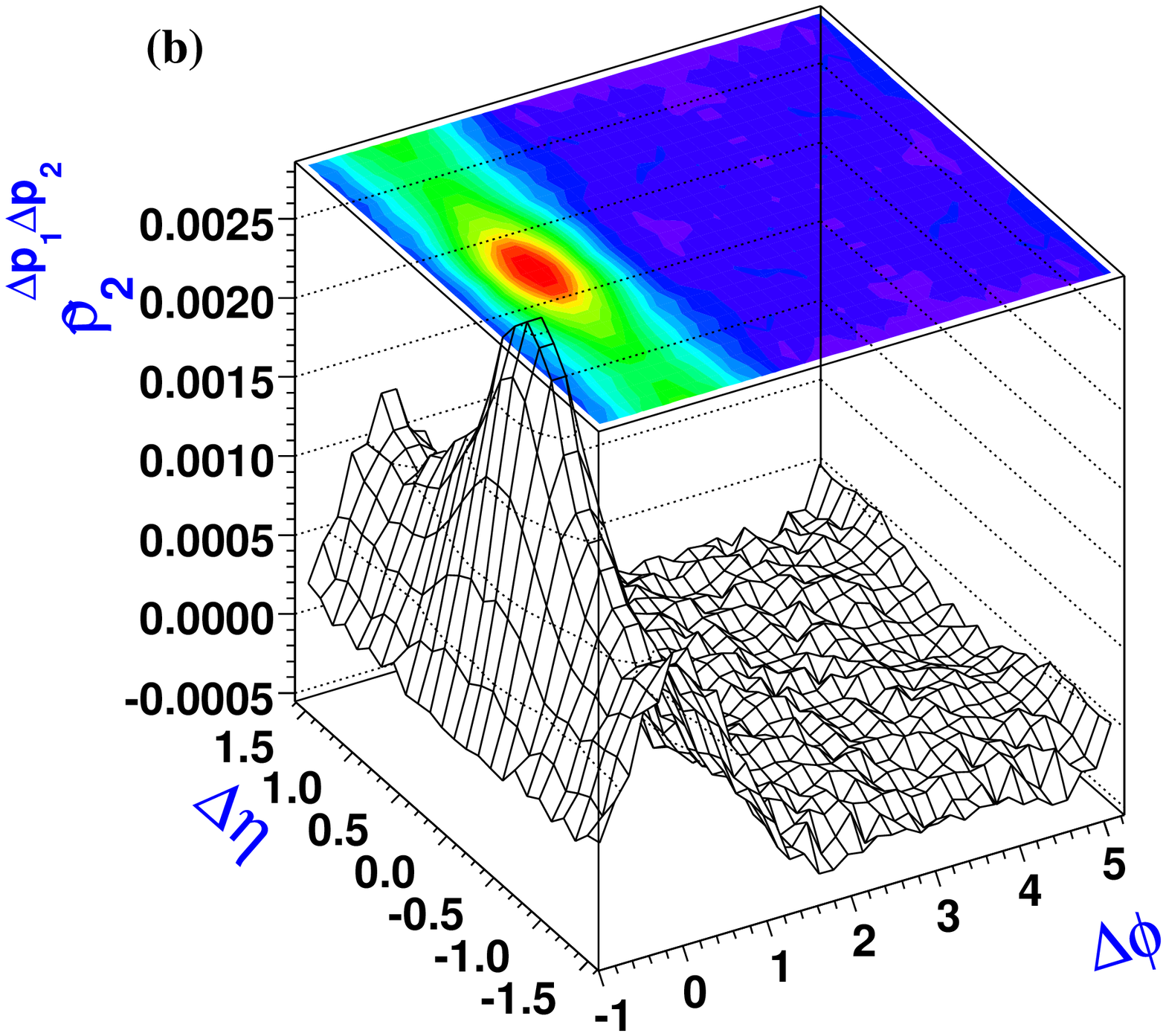}}
\resizebox{8.7cm}{6.5cm}{\includegraphics{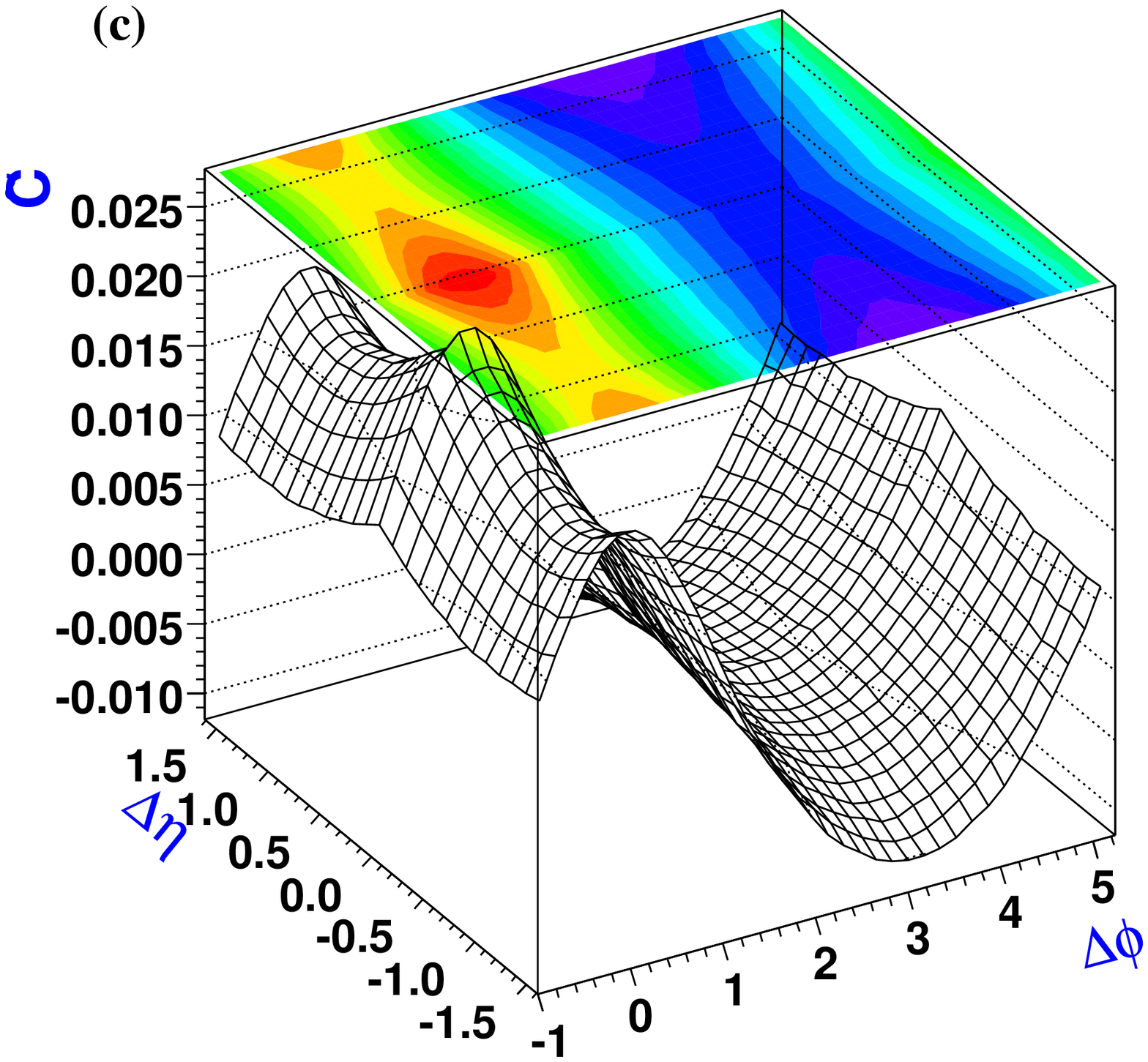}}
\caption[]{(Color online) Comparison of (a) {\it inclusive}, $\rho _{_2 }^{\Delta p_1 \Delta p_2 } \left( 
{\Delta \eta ,\Delta \varphi } \right)$, (b) {\it event-wise}, $\tilde \rho _{_2 }^{\Delta p_1 \Delta p_2 } 
\left( {\Delta \eta ,\Delta \varphi } \right)$ and (c) $\tilde C$ transverse momentum correlation functions 
obtained with radially boosted ($\beta = $ 0.3) {\it N} = 15 $p+p$ collisions at $\sqrt{s}=200$ GeV, 
generated with PYTHIA. }
\label{fig:4}
\end{figure}

We next consider the effect of radial flow on artificially generated 
$A+A$ collisions. As for Figures \ref{fig:4}(a-c), we produce $A+A$ collisions 
consisting of {\it N}~=~15 independent $p+p$ interactions. Particles of a given interaction are now boosted radially in the 
transverse plane with $\beta=0.3$. The dominant features remains the same as discussed previously for Figures \ref{fig:2}(a-c), i.e., 
observation of diminished away-side ridge-like correlations and enhanced near-side correlation for all the three observables. Therefore, 
on the basis of simulated $p+p$ and $A+A$ collisions, we conclude that strong radial flow should shift 
the away side structure to the near side in $A+A$ collisions and thereby produce a ridge in momentum correlations. 

\section{Effect of Flow on transverse momentum correlations in heavy ion collisions}
\label{Sect:FlowEffects}

Flow plays an important role in heavy ion collisions. It particularly affects particle correlations in azimuth. The 
effect of elliptic flow on azimuthal (density) correlations is well documented \cite{WhitePapers}. Here, we 
consider its effect on transverse momentum correlations studied as a function of relative azimuthal angles and 
pseudorapidity. The calculation method is similar to that used in \cite{Pruneau06}.

We model the probability density function (pdf) of particle emission as a function of pseudorapidity, azimuthal direction
 and transverse momentum, relative to reaction plane as follows.
\begin{equation}
P_1 (\eta ,\varphi ,p  |\Psi ) = P_1 (\eta, p )\left\{ {1 + 2\sum\limits_n {v_n (\eta,p )} \cos \left( 
{n\left( {\varphi  - \Psi } \right)} \right)} \right\}
\label{Eq42}
\end{equation}

$\Psi$ represents the reaction plane azimuthal direction. $\eta$, $\varphi$, and $p_t$ are the 
pseudorapidity, azimuthal emission angle and transverse momentum of the particle, respectively. $P_1 (\eta ,p)$ 
is the pdf of the particle at a given pseudorapidity and transverse momentum. Explicit knowledge of this pdf is 
not required in the following. Assuming the reaction plane is not measured, the two-particle pdf is given by
\begin{equation}
P_2 (\eta _1 ,\varphi _1 ,p_{1} ,\eta _2 ,\varphi _2 ,p_{2} ) = P_1 (\eta _1 ,p_{1} )P_1 (\eta _2 
,p_{2} ) \nonumber \\ 
\times \left\{ {1 + 2\sum\limits_n {v_n (\eta _1 ,p_{1} )v_n (\eta _2 ,p_{2} )} \cos \left( {n\left( 
{\varphi _1  - \varphi _2 } \right)} \right)} \right\} 
\label{Eq43}
\end{equation}

We use Eq. \ref{Eq13} to calculate the {\it inclusive} $\Delta p_{1} \Delta p_{2}$ correlations:
\begin{equation}
\rho _2^{\Delta p_{1} \Delta p_{2} } (\eta _1 ,\varphi _1 ,\eta _2 ,\varphi _2 ) = \frac{{2\sum\limits_n {\left( 
{v_n^{p} (\eta _1 ) - \langle p (\eta _1 )\rangle v_n (\eta _1 )} \right)\left( {v_n^{p } (\eta _2 ) - 
\langle p(\eta _2 )\rangle v_n (\eta _2 )} \right)\cos \left( {n\left( {\varphi _1  - \varphi _2 } \right)} \right)} }}{{1 + 
2\sum\limits_n {v_n (\eta _1 )v_n (\eta _2 )} \cos \left( {n\left( {\varphi _1  - \varphi _2 } \right)} \right)}}
\label{Eq44}
\end{equation}

where $v_n (\eta )$ and $v_n^{p_t } (\eta )$
are average and $p_t$ weighted average of flow coefficients function as a function of pseudorapidity, respectively.
\begin{eqnarray}
v_n (\eta ) & = & \frac{1}{{P_n (\eta )}}\int {P_n (\eta ,p )v_n (\eta ,p )dp }  \\ 
\label{Eq45}
v_n^{p_t } (\eta ) & = & \frac{1}{{P_n (\eta )}}\int {P_n (\eta ,p )v_n (\eta ,p )p dp  } 
\label{Eq46}
\end{eqnarray}
and 
\begin{equation}
\bar p(\eta ) = \frac{1}{{P_n (\eta )}}\int {P_n (\eta ,p )p dp } 
\end{equation}

We find the magnitude of the $\Delta p_t \Delta p_t$ correlations associated with a flowing medium are 
determined by the flow coefficients weighted by the transverse momentum and the averaged transverse 
momentum of the system. The correlations exhibit cosine modulations in azimuth. It is 
observed in $Au+Au$ collisions at RHIC that flow coefficients and average transverse momentum are both 
functions of the pseudorapidity of the particles \cite{phobos,brahms}. We therefore expect, based on Eq. \ref{Eq13}, 
the $\Delta p_t \Delta p_t$  two-particle correlations associated with flow are also a function of the particles 
pseudorapidity difference. Azimuthal anisotropy coefficients measured at RHIC are largely dominated by elliptic 
flow ($v_2$) coefficients. Flow modulations  of the {\it inclusive} correlations should then also dominated by second 
order harmonic coefficients.

\section{Experimental Robustness of Correlation Functions}
\label{Sect:Robustness}

Measurements of correlation observables in $p+p$ and $A+A$ collisions require one applies corrections for the 
limited acceptance and detection efficiency.  In some cases, correlation functions may be formulated in such a 
way that detection efficiency partly cancels. This was discussed already 
by a number of authors for various correlation observables. See for instance Ref. \cite{NystrandTydesjo} 
for a discussion of the robustness of factorial moments and the observable, $\nu_{+-,dyn}$, used 
in measurements of charge correlations. In this section, we investigate the extent to which the transverse 
momentum differential correlation functions, defined in Sect. \ref{Sect:PtCorrelations}, are robust observables. 
We use simple simulations to explicitly verify the robustness of these observables in contexts where efficiencies 
are non-trivial functions of the particle kinematics.

 Neglecting instrumental effects such as track splitting, ghost tracks {\it etc}., the single-particle and pair 
yields, $N_1 (x)$ and $N_2 (x_i ,x_j )$ measured at some $p_t$, $\eta$, $\varphi$,  are the products of the one-
 and two-particle densities by efficiencies $\varepsilon _1 (x)$ and $\varepsilon _2 (x_i ,x_j )$ for measuring 
one- and two- particles, respectively.

\begin{eqnarray}
N_1 (x) & = & \varepsilon _1 (x)\rho _1 (x) \\ 
N_2 (x_i ,x_j ) & = & \varepsilon _2 (x_i ,x_j )\rho _2 (x_i ,x_j ) 
\label{Eq:70}
\end{eqnarray}

In the above equations, $x$ represents the measured variables $p_t$, $\eta$, $\varphi$. The  `raw' momentum correlation, 
$\rho _{2,measured}^{\Delta p_1,\Delta p_2 } (\eta _1 ,\eta _2 )$, is thus a function of the detection efficiency. We show in the 
Appendix that particle 
losses due to limited efficiency in general lead to a modification of the fixed 
multiplicity momentum averages. Measurements of average momentum are consequently intrinsically non-robust. However, one 
also finds that if the average $p_t$ is independent of, or varies slowly with, the event multiplicity, 
one can apply Eq. \ref{Eq:70} to evaluate the correlation functions.
\begin{eqnarray}
\rho _{2,raw}^{\Delta p_1 \Delta p_2 } (\eta _1 ,\varphi _1 ,\eta _2 ,\varphi _2 ) = \frac{{\int {\varepsilon _2 (\eta _1 ,\varphi 
_1 ,p_{1} ,\eta _2 ,\varphi _2, p_{2} )\rho _2 (\eta _1 ,\varphi _1 ,p_{1} ,\eta _2 ,\varphi _2 
,p_{2} )p_{1} p_{2} dp_{1} dp_{2} } }}{{\int {\varepsilon _2 (\eta _1 ,\varphi _1 
,p_{1} ,\eta _2 ,\varphi _2 ,p_{2} )\rho _2 (\eta _1 ,\varphi _1 ,p_{1} ,\eta _2 ,\varphi _2 ,p_{2} )dp_{1} dp_{2} } }}
\label{Eq:80}
\end{eqnarray}

Obviously, if the efficiency is independent of the measurement coordinates, {\it i.e.} 
$ \varepsilon _2 (\eta _1 ,p_1 ,\varphi _1 ,\eta _2 ,p_2 ,\varphi _2 ) = {\rm{constant}}, $ 
then efficiencies cancel out in Eq. \ref{Eq:80}: the {\it inclusive} transverse momentum correlation is thus a 
``robust" quantity.  In general, however, the efficiency is a function of the measurement 
coordinates.  This implies corrections are required to account for detection efficiency. We note that Eq. \ref{Eq:80} 
simplifies considerably if the efficiency can be factorized into separate functions of momentum and pseudorapidity. It 
reduces to $\rho _2^{\Delta p_1, \Delta p_2 } (\eta _1 ,\eta _2 )$, {\it i.e.} is robust to first order, if the efficiency is uniform 
across the $p_t$ acceptance of the measurement.  Large 
detectors at RHIC and LHC have detection efficiencies that are reasonably uniform across their acceptance. 
However, various instrumental effects, such as detector boundaries or defective components, introduce small non-uniformities. 
We model these as follows:
	
\begin{eqnarray}
\varepsilon _1 (x_1 ) & = & \varepsilon _{1,0} \left( {1 + \delta _1 (x_1 )} \right) \\ 
\label{Eq:81}
\varepsilon _2 (x_1 ,x_2 ) & = & \varepsilon _{2,0} \left( {1 + \delta _2 (x_1 ,x_2 )} \right) 
\label{Eq:82}
\end{eqnarray}

where $\varepsilon_{1,0}$ and $\varepsilon_{2,0}$  correspond to the average efficiencies for detecting one- and 
two-particles, respectively. The functions $\delta_1(x_1)$ and $\delta_2(x_1,x_2)$  represent detection non-uniformities 
across the detector acceptance in $p_t$, $\varphi$ and $\eta$. They average to zero, by definition, over the 
acceptance of the apparatus. For ``reasonably behaved'' detectors, one typically finds
\begin{equation}
\varepsilon _{2,0}  \simeq \left( {\varepsilon _{1,0} } \right)^2  \nonumber
\end{equation}
Let us assume further that $\delta _i  \ll 1$.  One then finds the raw two-particle momentum correlation is approximately 
equal to
\begin{equation}
\rho _{2,raw}^{\Delta p_{1}\Delta p_{2}} (\Delta \eta ,\Delta \varphi )  \approx  \rho _2^{\Delta p_{1}\Delta p_{2}} (\Delta \eta 
,\Delta \varphi )\left\{ {1 + a_2 (\Delta \eta ,\Delta \varphi ) - b_2 (\Delta \eta ,\Delta \varphi )} \right\}
\label{Eq:84}
\end{equation}
where
\begin{eqnarray}
a_2 (\Delta \eta ,\Delta \varphi ) & = & \frac{{\int {\rho _2 (\eta _1 ,\eta _2 ,\varphi _1 ,\varphi _2 ,p_1 ,p_2 )\delta _2 
(\eta _1 ,\eta _2 ,\varphi _1 ,\varphi _2 ,p_1 ,p_2 )\Delta p_1 \Delta p_2 dp_1 dp_2 } }}{{\int {\rho _2 (\eta _1 
,\eta _2 ,\varphi _1 ,\varphi _2 ,p_1 ,p_2 )dp_1 dp_2 } }} \\
b_2 (\Delta \eta ,\Delta \varphi ) & = & \frac{{\int {\rho _2 (\eta _1 ,\eta _2 ,\varphi _1 ,\varphi _2 ,p_1 ,p_2 )\delta _2 
(\eta _1 ,\eta _2 ,\varphi _1 ,\varphi _2 ,p_1 ,p_2 )dp_1 dp_2 } }}{{\int {\rho _2 (\eta _1 ,\eta _2 ,\varphi _1 
,\varphi _2 ,p_1 ,p_2 )dp_1 dp_2 } }}
\end{eqnarray}

In general one has $\left| {\Delta p_1 \Delta p_2 } \right| < 1$, which implies $a_2  < b_2$, {\it i.e.} the difference $a_2  - b_2$
 is non-vanishing. The {\it inclusive} transverse momentum correlation is indeed not perfectly robust as an observable. 
This implies a correction might be explicitly needed. A correction based on Eq. \ref{Eq:81} is not practical because it 
requires knowledge of the two-particle density. It is thus simpler to carry an efficiency correction  on a per 
particle pair basis, {\it i.e.} rather than counting ``1" for each measured pair, increment the number of pairs according 
to
\begin{equation}
1/\varepsilon _2 (\eta _1 ,\eta _2 ,\varphi _1 ,\varphi _2 ,p_1 ,p_2 ) \nonumber
\end{equation}
The above formulation of efficiency effects on the correlation neglects possible event detection biases incurred with small 
detection efficiency. Large particle losses may produce a significant modification of the 
$p_t$ spectrum of small multiplicity events. This is, thus, susceptible of significantly biasing the measurement of 
$p_t$ correlation of $p+p$ collisions.

\begin{figure}[!htp]
\centering
\resizebox{8.7cm}{6.5cm}{\includegraphics{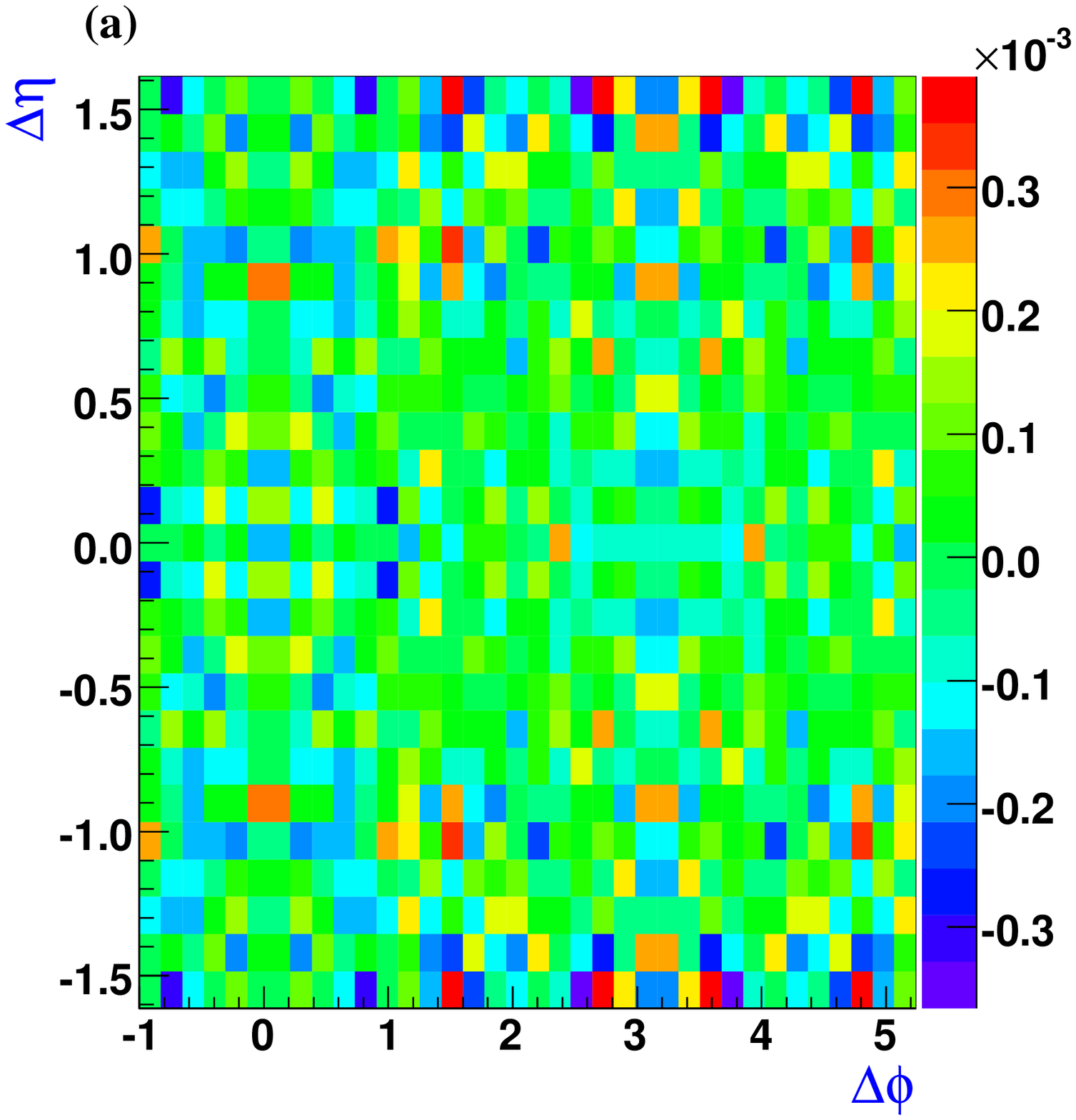}}
\resizebox{8.7cm}{6.5cm}{\includegraphics{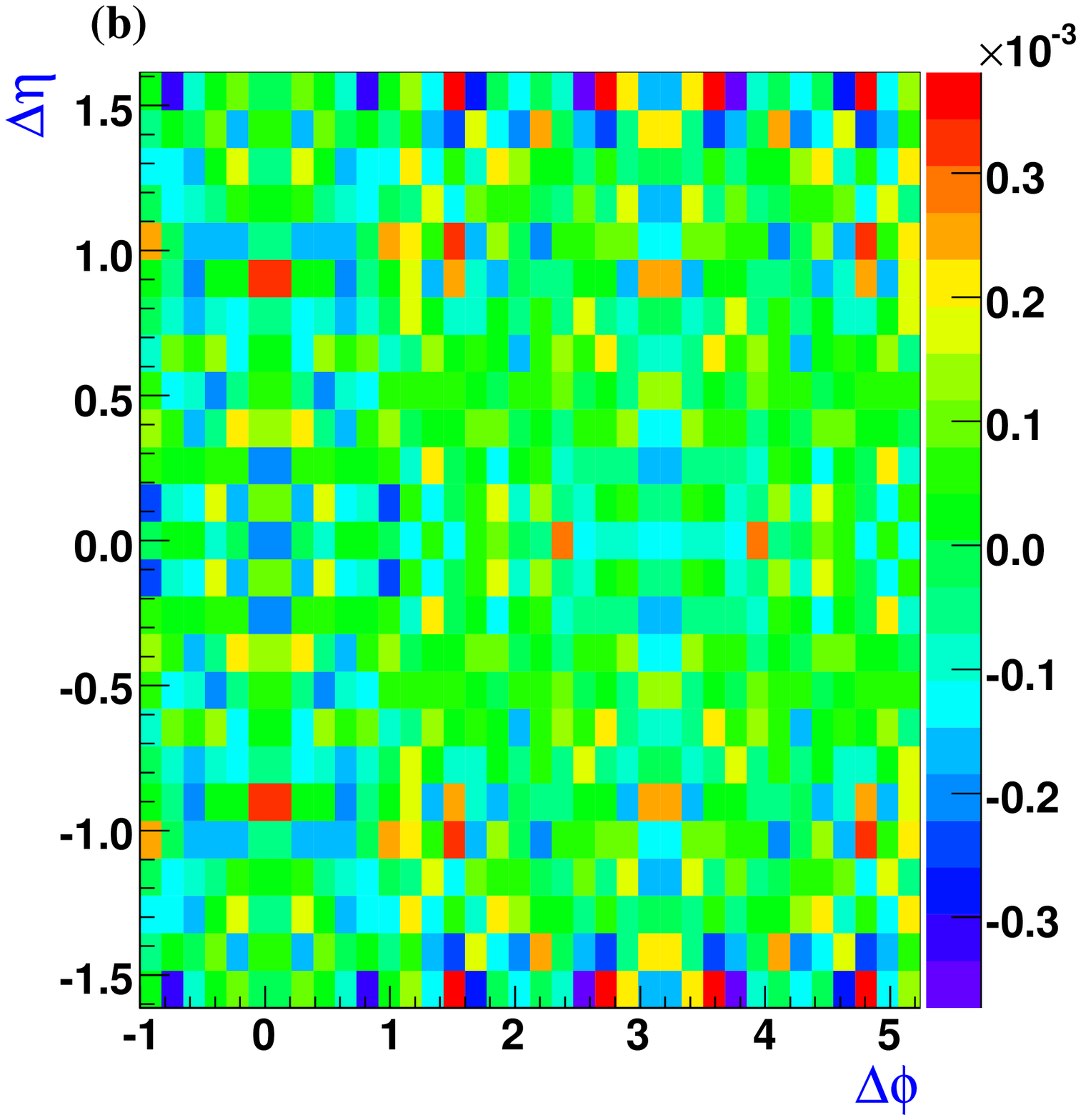}}
\resizebox{8.7cm}{6.5cm}{\includegraphics{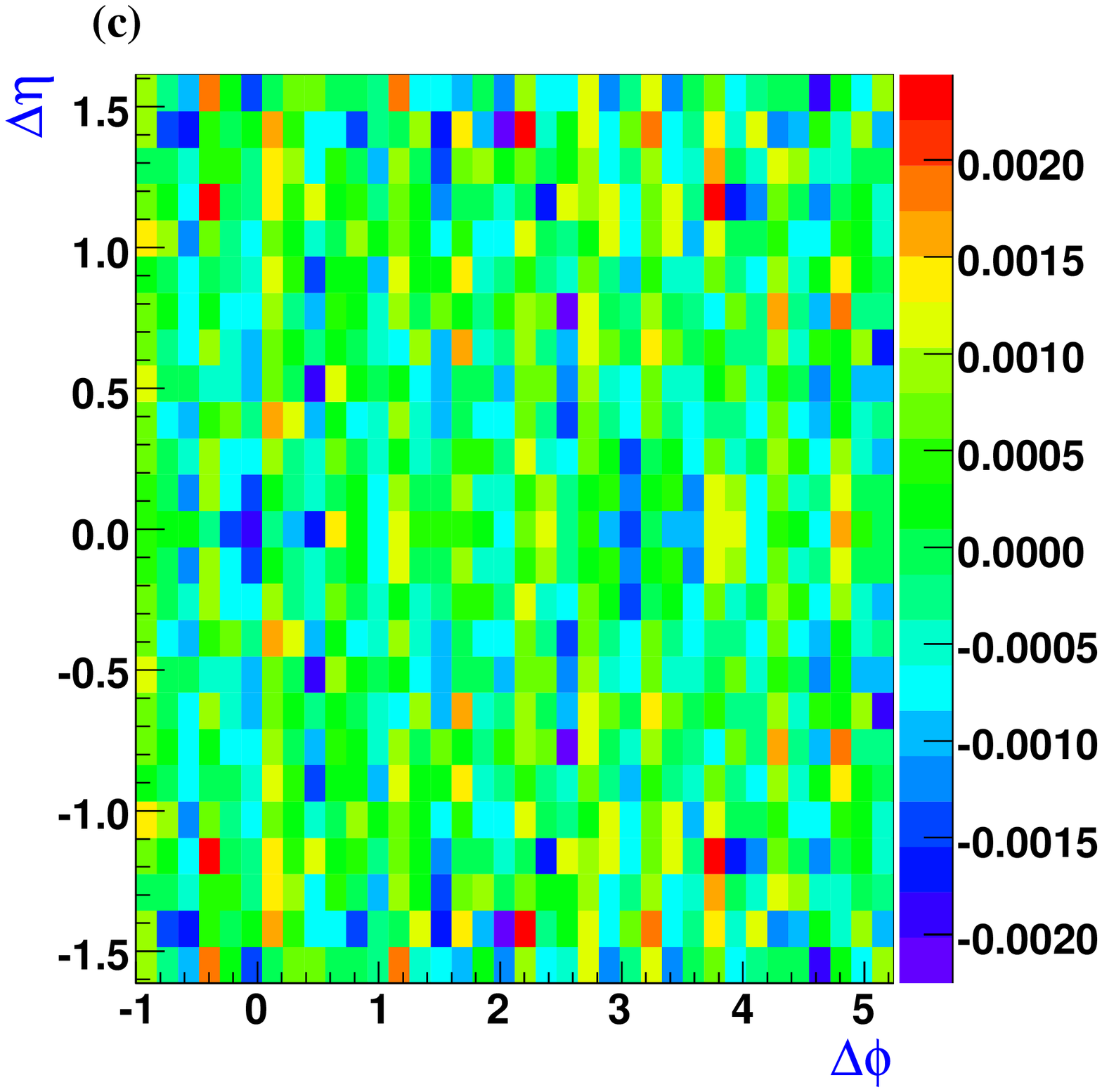}}
\caption[]{(Color online) Difference of the {\it inclusive}, {\it event-wise} and $\tilde C$ transverse momentum correlations 
function obtained with $p+p$ collisions at $\sqrt{s}=200$ GeV generated with PYTHIA for perfect efficiency, $\varepsilon = 1$ and
reduced efficiency, $\varepsilon = 0.8$. (a) {\it inclusive} average 
$\rho _{_2,\varepsilon = 1 }^{\Delta p_1 \Delta p_2 } \left( {\Delta \eta ,\Delta \varphi } \right) - \rho _{_2,\varepsilon = 0.8 }^{\Delta p_1 \Delta p_2 } 
\left( {\Delta \eta ,\Delta \varphi } \right)$, (b) {\it event-wise} average 
$\tilde \rho _{_2,\varepsilon = 1 }^{\Delta p_1 \Delta p_2 } \left( {\Delta \eta ,\Delta \varphi } \right) - \tilde \rho _{_2,\varepsilon = 0.8 }^{\Delta p_1 \Delta p_2 } \left( {\Delta \eta ,\Delta \varphi } \right)$ and (c) observable $\tilde C$, $\tilde C_{\varepsilon = 1} - \tilde C_{\varepsilon = 0.8}$. }
\label{fig:6}
\end{figure}

\begin{figure}[!htp]
\centering
\resizebox{8.7cm}{6.5cm}{\includegraphics{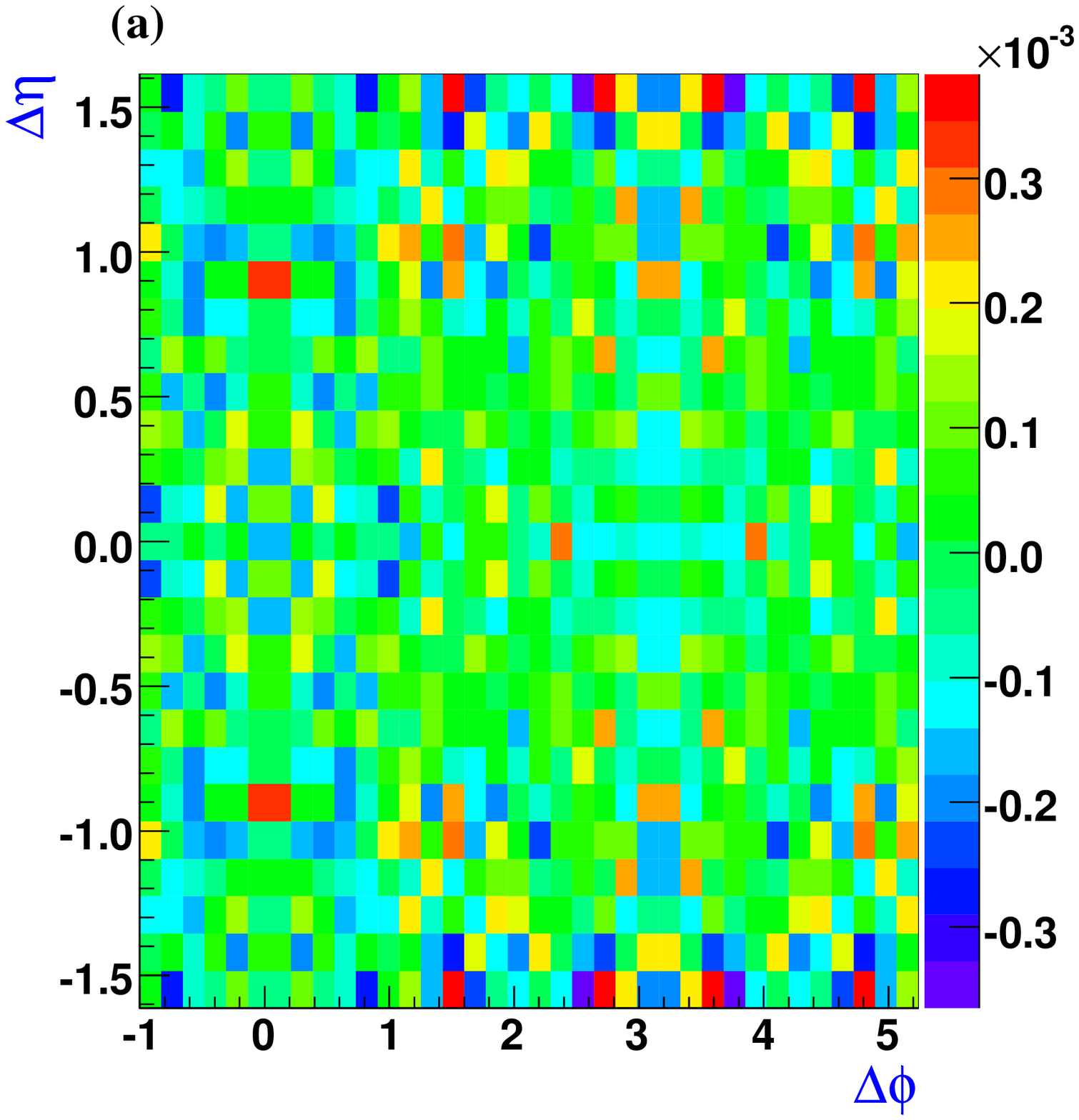}}
\resizebox{8.7cm}{6.5cm}{\includegraphics{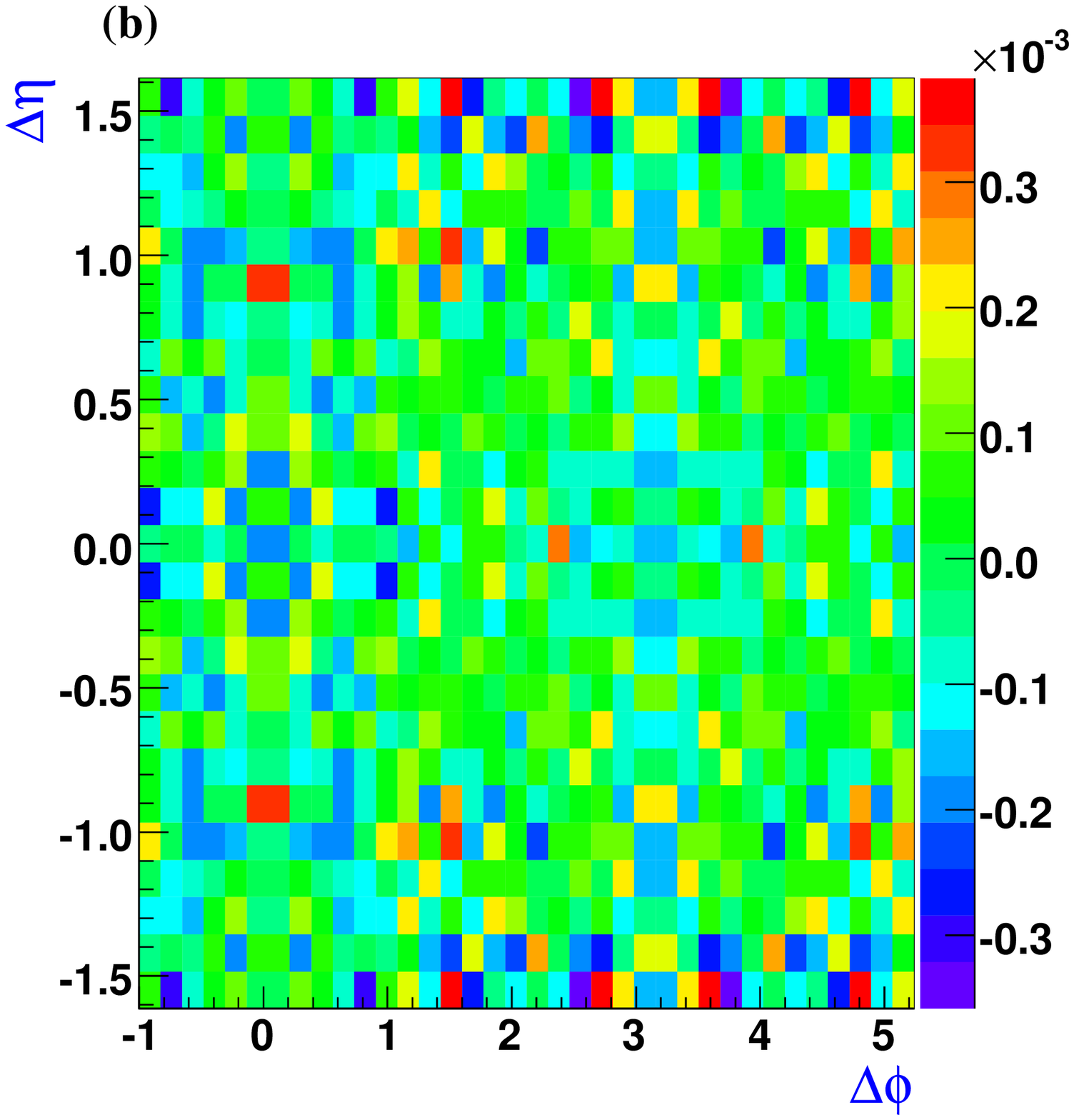}}
\resizebox{8.7cm}{6.5cm}{\includegraphics{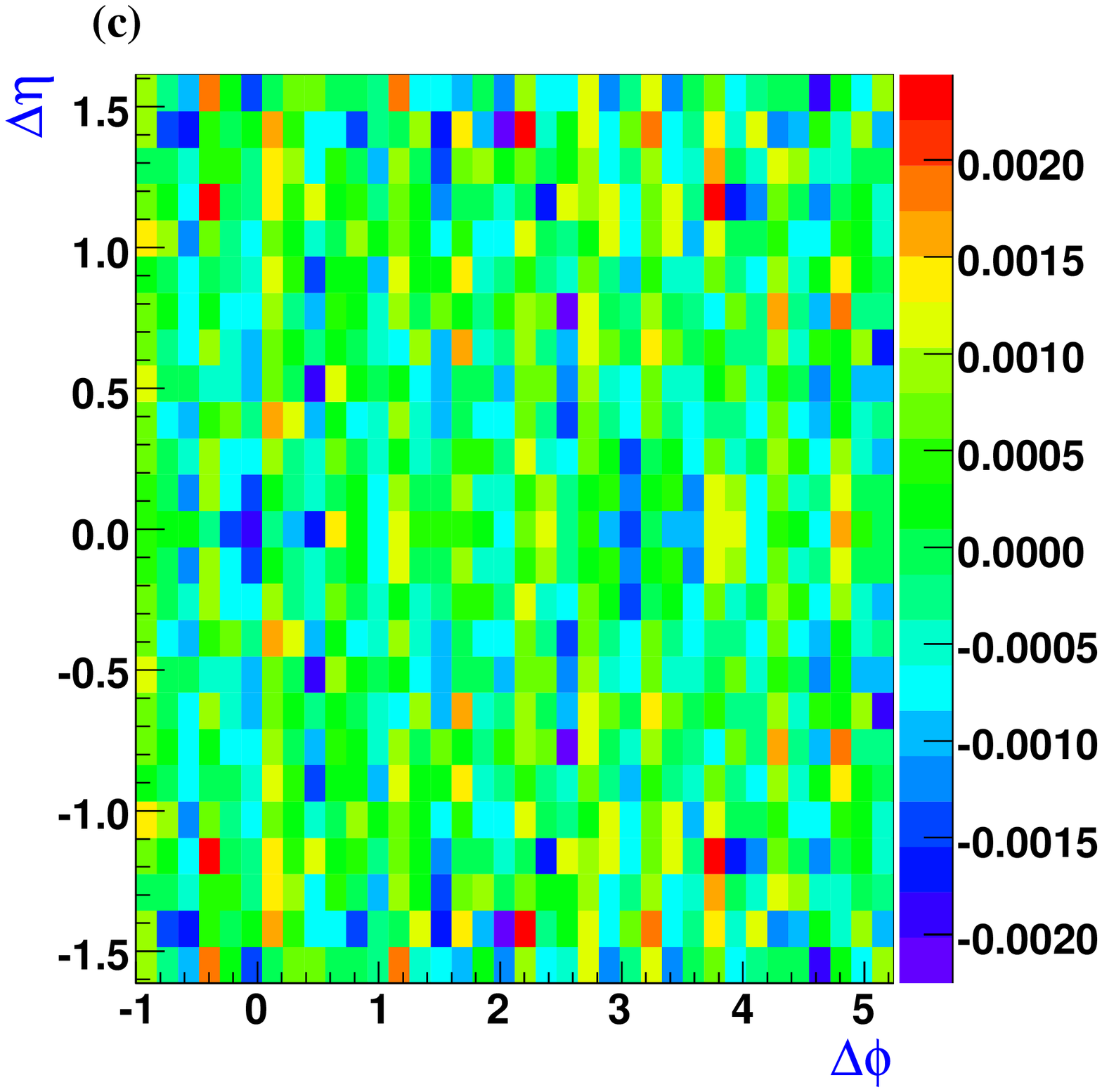}}
\caption[]{(Color online) Difference of the {\it inclusive}, {\it event-wise} and $\tilde C$ transverse momentum correlations 
function obtained with $p+p$ collisions at $\sqrt{s}=200$ GeV generated with PYTHIA for perfect efficiency, $\varepsilon = 1$ and
efficiency dependent on particle transverse momentum as per Eq. \ref{Eq:85} using $\varepsilon_{0}$ = 0.8 and a = 0.05 (see Eq. \ref{Eq:85}). 
(a) {\it Inclusive} average 
$\rho _{_2,\varepsilon = 1 }^{\Delta p_1 \Delta p_2 } \left( {\Delta \eta ,\Delta \varphi } \right) - \rho _{_2,\varepsilon = 0.8 }^{\Delta p_1 \Delta p_2 } 
\left( {\Delta \eta ,\Delta \varphi } \right)$, (b) {\it event-wise} average 
$\tilde \rho _{_2,\varepsilon = 1 }^{\Delta p_1 \Delta p_2 } \left( {\Delta \eta ,\Delta \varphi } \right) - \tilde \rho _{_2,\varepsilon = 0.8 }^{\Delta p_1 \Delta p_2 } \left( {\Delta \eta ,\Delta \varphi } \right)$ and (c) observable $\tilde C$, $\tilde C_{\varepsilon = 1} - \tilde C_{\varepsilon = 0.8}$. }
\label{fig:7}
\end{figure}

Figure \ref{fig:1}, discussed in Sect. \ref{Sect:Model}, displays PYTHIA events obtained with perfect efficiency
and serves as a reference for simulations presented in Figures \ref{fig:6} through \ref{fig:7} obtained with limited detection efficiency. 
The effect of detection efficiency is obtained by individually and randomly rejecting particles on the basis of a given 
efficiency prescription, 
\begin{equation}
\varepsilon _2  = f(\eta _1 ,\eta _2 ,\varphi _1 ,\varphi _2 ,p_1 ,p_2 ) \nonumber
\end{equation}

\begin{figure}[!htp]
\centering
\resizebox{8.7cm}{6.5cm}{\includegraphics{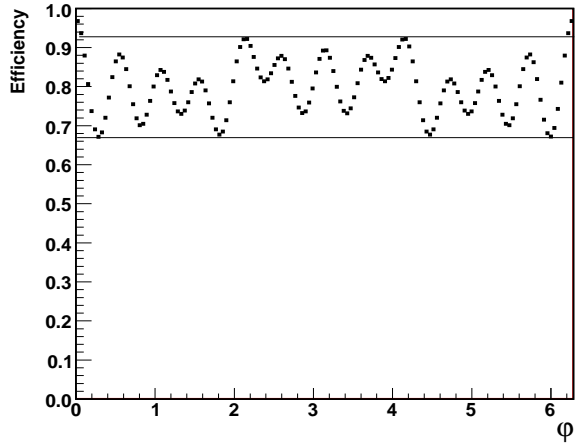}}
\caption[]{Efficiency non-uniformities are shown as a function of azimuthal angle, $\varphi$. See text for details. }
\label{fig:5}
\end{figure}

One uses a two-particle efficiency equal to the 
product of single particle efficiencies, as per Eq. \ref{Eq:82}. The single particle efficiency is defined on the 
basis of a Fourier decomposition involving twelve terms as follows:
\begin{equation}
\varepsilon (\varphi ) = \varepsilon _0 \left( {1 + \sum\limits_{n = 1}^{12} {\varepsilon _i \cos (n\varphi )} } 
\right)
\label{Eq:83}
\end{equation}

The coefficients $\varepsilon _i $ are chosen to approximately model the azimuthal efficiency dependency of large 
detectors such as the STAR TPC and ALICE TPC. Spike structures are introduced in azimuth 
to simulate the effect of detector boundaries. Twelve sectors are used. The amplitude of the 
efficiency varies across the acceptance by $\pm$10\%. Figure \ref{fig:5} shows efficiency non-uniformities introduced in azimuth. 
Figures \ref{fig:6} shows the robustness of the observables studied on the basis of Eq. \ref{Eq:83}.
 Our study of the impact of detection efficiency is based on single $p+p$ collisions generated with PYTHIA.

Figures \ref{fig:6}(a-c) show correlations obtained with Eq. \ref{Eq:83} using $\varepsilon_{0} = 0.8$.
For each produced particle one generates a random number, $0<r<1$. The particle is included 
in the analysis provided $r<\varepsilon$. 
In Figures \ref{fig:6}(a-c) we show the difference between the correlation functions obtained for 
perfect efficiency, {\it i.e.} $\varepsilon = 1$ and reduced efficiency, {\it i.e.} $\langle \varepsilon \rangle = 0.8$, 
for all three observables of interest. 
The differences in the case of {\it inclusive} and {\it event-wise} observables are 
of the order of 0.5\% only, whereas the difference observed for $\tilde C$ is 1\%. However, despite extremely small dependence 
on efficiency, we observe finite near-side and away-side structures in the difference. In contrast, $\tilde C$ reflects the 
effect of detector boundaries. There are exactly twelve structures in azimuth which correspond to detector 
boundaries introduced as non-uniformities in azimuth. The differences are however numerically small and amount to $\sim$1\% of the 
signal at $\Delta \varphi$ = $\Delta \eta$ = 0. We therefore 
conclude that all the three observables exhibit very little dependence on detection efficiency.  

We next include, for illustrative purposes, a linear efficiency dependency on the particle transverse momentum.
\begin{equation}
\varepsilon (\varphi ,p) = \varepsilon _0 \left( {1 - ap} \right)\left( {1 + \sum\limits_{n = 1}^{12} {\varepsilon _i 
\cos (n\varphi )} } \right)
\label{Eq:85}
\end{equation}

Figures \ref{fig:7}(a-c) show the difference between the correlation functions obtained for perfect efficiency, {\it i.e.} $\varepsilon$ = 1 and 
efficiency dependent on particle transverse momentum as per Eq. \ref{Eq:85} obtained with $\varepsilon_{0}$ = 0.8. 
The dominant features observed in Figures \ref{fig:7}(a-c) remain the same as those observed in Figures \ref{fig:6}(a-c). 
We verified the robustness of these observables 
by further reducing the efficiency (Eq. \ref{Eq:82}) to 50\%, {\it i.e.} $\varepsilon_{0}$ = 0.5. The difference, 
$\rho _{_2,\varepsilon = 1 }^{\Delta p_1 \Delta p_2 } \left( {\Delta \eta ,\Delta \varphi } \right) - \rho _{_2,\varepsilon = 0.5 }^{\Delta p_1 \Delta p_2 } \left( {\Delta \eta ,\Delta \varphi } \right)$, is of the order of 5\% for {\it inclusive} and {\it event-wise} observables where as the difference for $\tilde C$ 
increases by 2\% only. 
We conclude that all three observables, {\it inclusive}, {\it event-wise} and $\tilde C$ are quite robust. Their dependence on efficiency, even when it 
varies across the acceptance by as much as by 10\%, is very small.  

\section{Summary and Conclusions}
\label{Sect:Summary}

We introduced formal definitions and notations for {\it inclusive}, {\it event-wise} and $\tilde C$ correlation functions. 
We considered more specifically transverse momentum two-particle correlations expressed 
as functions of the particle pseudorapidity and azimuthal angle differences. We also studied a generalization of 
the observable $\tilde C$ proposed by Gavin \cite{Gavin}, which he uses to determine the viscosity per unit of entropy of the 
nuclear matter produced in $A+A$ collisions at RHIC and the  LHC. 

We used the event generator PYTHIA \cite{Pythia} to study and predict the shape of the correlation functions 
for minimum bias $p+p$ collisions at $ \sqrt{s}=200$ GeV. Simulations were carried out for particles in the 
ranges $|\eta|<1$ and $0.2<p_t<2.0$ GeV/c. The correlation functions exhibit a relatively narrow peak at $\Delta 
\eta = \Delta \varphi = 0$, and a wide away-side band or ridge extending over the full range of pseudorapidity 
$|\Delta \eta| < 2$. These features are qualitatively similar to those found in particle density correlations 
calculated with PYTHIA \cite{Pruneau07}.

While the three observables are meant to determine the level of transverse momentum correlations between 
produced particles, their definitions differ significantly. One expects in the large multiplicity limit, values 
obtained with the three observables should be identical. However, in practice, multiplicities in a given bin of  
relative pseudorapidity and azimuthal angles ($\Delta \eta, \Delta \varphi$) are rather small. This 
implies differences between these three observables can be substantial.  We verified this expectation on the basis 
of $p+p$ collision simulations calculated with PYTHIA event generator \cite{Pythia}. We observed {\it inclusive} and {\it event-wise} 
observables to be approximately identical. Both, however, differ significantly from $\tilde C$. 
We conclude that for a precise comparative study of 
transverse momentum correlation functions for different colliding systems, at different energies and from different experiments, 
it is essential the same observable be used to characterize the correlation shape and strength.

We studied the experimental robustness of the three observables. We studied the impact of detection 
efficiency using PYTHIA simulated $p+p$ collision events with an arbitrary parameterization of the efficiency 
dependence on the detection angle and transverse momentum. We found that these observables are robust even when the 
detection efficiency reduces to 50\%. 

We studied the scaling of the {\it inclusive} transverse momentum correlation function with the number of 
participating nucleons in $A+A$ collisions assuming the $p+p$ interactions are independent and with no 
rescattering of the secondaries. We found that, similar to other correlation observables \cite{Pruneau03}, the 
correlation strength exhibits a $1/N_{part}$ dependence for varying $A+A$ collision centralities. 

We additionally studied the effect of radial flow in $p+p$ and $A+A$ collisions on the magnitude and 
azimuthal dependence of these observables. We found that radial flow is responsible for the shift of the away side structure 
to the near side ridge like structure.

{\bf Acknowledgements}\\
The authors acknowledge constructive discussions with S. Gavin and S. Voloshin. The authors also
thank L. Tarini for his careful reading of the manuscript and editorial comments. 
This work was supported in part by DOE grant no. DE-FG02-92ER40713.

\section{Appendix}
Experimentally, one measures at a given $p_t$, $\eta$, and $\varphi$ a number of particles $N(p_t,\eta,\varphi)$. 
This number is divided by the bin size to obtain the uncorrected density $\rho_{exp}(p_t,\eta,\varphi)$. 
In general, the detection efficiency is a function of all three coordinates herein noted $\epsilon(p_t$, $\eta$, $\varphi)$. 
The actual density, $\rho_{TH}(p_{t},\eta,\varphi)$, is obtained as follows:
\begin{equation}
\rho _{TH} (p_t ,\eta ,\varphi ) = \frac{{\rho _{\exp } (p_t ,\eta ,\varphi )}}{{\varepsilon (p_t ,\eta ,\varphi )}}
\end{equation}
The efficiency changes as detector conditions evolve over time. It may also be a complicated function of the measured event structure. 
It is the case for instance with a TPC in heavy ion collisions measurements. One observes experimentally 
that the detection efficiency depends on the detector occupancy, or particle track multiplicity in the detector.  At low multiplicity,
 tracks are on average well separated and hence easy to identify and reconstruct. However, for large multiplicity, the distance between tracks 
is reduced such that they may cross and partially overlap. The efficiency is consequently a function of the event multiplicity as well 
as the beam luminosity.  The dependence on luminosity may be corrected for by estimating $\epsilon(p_t$, $\eta$, $\varphi)$ based 
on embedding techniques. We parameterize this dependence on global event factors with an index $m$. 
For convenience, one writes the detection efficiency at given multiplicity as follows:
\begin{equation}
\varepsilon (p_t ,\eta ,\varphi |m) = \varepsilon _0 (\eta ,\varphi |m)B(p_t ,\eta ,\varphi |m)
\end{equation}
where $\varepsilon _0 (p_t ,\eta ,\varphi |m)$ represents the $p_t$ averaged efficiency, and B is a  normalized response function.
\begin{equation}
\int {B(p_t ,\eta ,\varphi |m)} dp_t  = 1
\end{equation}
The number of particles detected at given  $\eta$ and $\varphi$ is written as
\begin{equation}
N(p_t ,\eta ,\varphi ) = \sum\limits_{m}^{} {\varepsilon _0 (\eta ,\varphi |m)B(p_t ,\eta ,\varphi |m)\rho _{TH} (p_t ,\eta ,\varphi |m)} 
\end{equation}
The measured average particle $p_t$ is then
\begin{equation}
\left\langle {p_t } \right\rangle _{\exp } (\eta ,\varphi ) = \frac{{\sum\limits_{m} {\varepsilon _0 (\eta ,\varphi |m)
\int {B(p_t ,\eta ,\varphi |m)\rho _{TH} (p_t ,\eta ,\varphi |m)p_t dp_t } } }}{{\sum\limits_{m} {\varepsilon _0 (\eta ,\varphi |m)
\int {B(p_t ,\eta ,\varphi |m)\rho _{TH} (p_t ,\eta ,\varphi |m)dp_t } } }}
\end{equation}
and is clearly not a robust quantity. However, in cases where the average $p_t$ is measured over a small range of multiplicity, one has 
\begin{equation}
\left\langle {p_t } \right\rangle _{\exp } (\eta ,\varphi ) = \frac{{\int {B(p_t ,\eta ,\varphi )\rho _{TH} 
(p_t ,\eta ,\varphi )p_t dp_t } }}{{\int {B(p_t ,\eta ,\varphi )\rho _{TH} (p_t ,\eta ,\varphi )dp_t } }}
\end{equation}
while this expression indicates the measured average $p_t$ still depends on the $p_t$  detector response, 
it is easy to verify that in cases where the $p_t$ dependence is ÔmildÕ, the above expression is a reasonable 
approximation of the actual average momentum. Consider for instance a case where the density has an exponential 
dependence on the momentum 
$\rho  \propto \exp ( - p_t /T)$
where $T$ is a slope parameter describing the particle distribution. We evaluate the impact of the detector response 
using a linear approximation $B \propto 1 + ap_t $. Assume T has a value of 0.6. The average $p_t$ changes by 1\% relative 
to the actual average $p_{t}$  when the 
value of the response parameter $a$ is chosen to be 0.05. Smaller values of $a$ lead to smaller deviations.
One thus finds that unless $a$ is exceedingly large, the detector response has a rather limited 
impact, a few percent only, on the measured  average momentum.  It is, nonetheless, a function of the detection coordinates 
$\eta$ and $\varphi$: deviations of measured $\langle p_{t} \rangle$ may depend on $\eta$ and $\phi$.   
The above argument is straightforwardly repeated for measurements of correlations $\left\langle {p_t p_t } \right\rangle (\Delta\eta ,\Delta\varphi)$.
Given this quantity depends essentially on the square of the detector response, one expects the impact of a $p_t$ 
dependence should be rather small - although potentially a function of the particle coordinates  $\eta_1$, $\varphi_1$, $\eta_2$ and $\varphi_2$.
\end{document}